\documentclass{pasa}
\usepackage{graphicx}
\usepackage{amsmath}
\usepackage{amssymb}
\usepackage{times}
\usepackage{color}
\usepackage[usenames,dvipsnames,svgnames,table]{xcolor}
\usepackage{subcaption}

\definecolor{customhdrcolor}{rgb}{0.0,0.0,0.0}
\definecolor{customcitecolor}{rgb}{0.0,0.5,0.75}
\definecolor{customlinkcolor}{rgb}{0.0,0.5,0.75}

\usepackage[colorlinks=true,linkcolor=customlinkcolor,urlcolor=customlinkcolor,citecolor=customcitecolor,pdftex]{hyperref}

\ifpdf\else\usepackage{graphics}\fi
\usepackage{graphics}

\DeclareRobustCommand{\TUSSEN}[3]{#2}

\title[The MWA radio environment]{The low-frequency environment of the Murchison Widefield Array: radio-frequency interference analysis and mitigation}

\def\ANU{$^{1}$}
\def\CAASTRO{$^{2}$}
\def\ASTRON{$^{3}$}
\def\Curtin{$^{4}$}
\def\UWisc{$^{5}$}
\def\UW{$^{6}$}
\def\CASS{$^{7}$}
\def\SKASA{$^{8}$}
\def\Rhodes{$^{9}$}
\def\CfA{$^{10}$}
\def\ASU{$^{11}$}
\def\USydney{$^{12}$}
\def\Haystack{$^{13}$}
\def\RRI{$^{14}$}
\def\MIT{$^{15}$}
\def\UWA{$^{16}$}
\def\Victoria{$^{17}$}
\def\UMelbourne{$^{18}$}
\def\Tata{$^{19}$}

\author[A.~R.~Offringa et al.]{A.~R.~Offringa\ANU$^,$\CAASTRO$^,$\ASTRON\thanks{Corresponding author. E-mail: \url{offringa@gmail.com}},
R.~B.~Wayth\Curtin$^,$\CAASTRO,
N.~Hurley-Walker\Curtin,
D.~L.~Kaplan\UWisc,
N.~Barry\UW,
A.~P.~Beardsley\UW,
M.~E.~Bell\CASS$^,$\CAASTRO,
G.~Bernardi\SKASA$^,$\Rhodes$^,$\CfA,
J.~D.~Bowman\ASU, 
F.~Briggs\ANU$^,$\CAASTRO,
J.~R.~Callingham\USydney$^,$\CAASTRO,
R.~J.~Cappallo\Haystack, 
P.~Carroll\UW,
A.~A.~Deshpande\RRI, 
J.~S.~Dillon\MIT,
K.~S.~Dwarakanath\RRI,
A.~Ewall-Wice\MIT,
L.~Feng\MIT,
B.-Q.~For\UWA,
B.~M.~Gaensler\USydney$^,$\CAASTRO, 
L.~J.~Greenhill\CfA,
P.~Hancock\Curtin$^,$\CAASTRO,
B.~J.~Hazelton\UW,
J.~N.~Hewitt\MIT,
L.~Hindson\Victoria,
D.~C.~Jacobs\ASU, 
M.~Johnston-Hollitt\Victoria,
A.~D.~Kapi\'{n}ska\UWA$^,$\CAASTRO,
H.-S.~Kim\UMelbourne$^,$\CAASTRO,
P.~Kittiwisit\ASU,
E.~Lenc\USydney$^,$\CAASTRO,
J.~Line\UMelbourne$^,$\CAASTRO,
A.~Loeb\CfA,
C.~J.~Lonsdale\Haystack, 
B.~McKinley\ANU$^,$\CAASTRO$^,$\UMelbourne,
S.~R.~McWhirter\Haystack,
D.~A.~Mitchell\CASS$^,$\CAASTRO, 
M.~F.~Morales\UW, 
E.~Morgan\MIT, 
J.~Morgan\Curtin,
A.~R.~Neben\MIT,
D.~Oberoi\Tata, 
S.~M.~Ord\Curtin$^,$\CAASTRO,
S.~Paul\RRI,
B.~Pindor\UMelbourne,
J.~C.~Pober\UW,
T.~Prabu\RRI, 
P.~Procopio\UMelbourne$^,$\CAASTRO,
J.~Riding\UMelbourne$^,$\CAASTRO,
N.~Udaya~Shankar\RRI, 
S.~Sethi\RRI,
K.~S.~Srivani\RRI,
L.~Staveley-Smith\UWA$^,$\CAASTRO,
R.~Subrahmanyan\RRI$^,$\CAASTRO, 
I.~S.~Sullivan\UW,
M.~Tegmark\MIT,
N.~Thyagarajan\ASU,
S.~J.~Tingay\Curtin$^,$\CAASTRO, 
C.~M.~Trott\Curtin$^,$\CAASTRO,
R.~L.~Webster\UMelbourne$^,$\CAASTRO, 
A.~Williams\Curtin, 
C.~L.~Williams\MIT,
C.~Wu\UWA,
J.~S.~Wyithe\UMelbourne$^,$\CAASTRO,
Q. Zheng\Victoria
\\
\ANU{}Research School of Astronomy and Astrophysics, Australian National University, Canberra, ACT 2611, Australia \\
\CAASTRO{}ARC Centre of Excellence for All-sky Astrophysics (CAASTRO) \\
\ASTRON{}Netherlands Institute for Radio Astronomy (ASTRON), PO Box 2, 7990 AA Dwingeloo, The Netherlands\\
\Curtin{}International Centre for Radio Astronomy Research, Curtin University, Bentley, WA 6102, Australia\\
\UWisc{}Department of Physics, University of Wisconsin--Milwaukee, Milwaukee, WI 53201, USA\\
\UW{}Department of Physics, University of Washington, Seattle, WA 98195, USA\\
\CASS{}CSIRO Astronomy and Space Science, Marsfield, NSW 2122, Australia\\
\SKASA{}SKA SA, 3rd Floor, The Park, Park Road, Pinelands, 7405, South Africa\\
\Rhodes{}Department of Physics and Electronics, Rhodes University, PO Box 94, Grahamstown, 6140, South Africa\\
\CfA{}Harvard-Smithsonian Center for Astrophysics, Cambridge, MA 02138, USA\\
\ASU{}School of Earth and Space Exploration, Arizona State University, Tempe, AZ 85287, USA\\
\USydney{}Sydney Institute for Astronomy, School of Physics, University of Sydney, NSW 2006, Australia\\
\Haystack{}MIT Haystack Observatory, Westford, MA 01886, USA\\
\RRI{}Raman Research Institute, Bangalore 560080, India\\
\MIT{}Kavli Institute for Astrophysics and Space Research, Massachusetts Institute of Technology, Cambridge, MA 02139, USA\\
\UWA{}International Centre for Radio Astronomy Research, University of Western Australia, Crawley, WA 6009, Perth, Australia\\
\Victoria{}School of Chemical \& Physical Sciences, Victoria University of Wellington, Wellington 6140, New Zealand\\
\UMelbourne{}School of Physics, University of Melbourne, Parkville, VIC 3010, Australia\\
\Tata{}National Centre for Radio Astrophysics, Tata Institute for Fundamental Research, Pune 411007, India
}
\jid{PASA}
\doi{10.1017/pas.\the\year.xxx}
\jyear{\the\year}
\usepackage[authoryear]{natbib}
\bibpunct{(}{)}{;}{a}{}{,}
\begin{document}
\begin{abstract}
The Murchison Widefield Array (MWA) is a new low-frequency interferometric radio telescope built in Western Australia at one of the locations of the future Square Kilometre Array (SKA). We describe the automated radio-frequency interference (RFI) detection strategy implemented for the MWA, which is based on the \textsc{aoflagger} platform, and present 72--231-MHz RFI statistics from 10 observing nights. RFI detection removes 1.1\% of the data. RFI from digital TV (DTV) is observed 3\% of the time due to occasional ionospheric or atmospheric propagation. After RFI detection and excision, almost all data can be calibrated and imaged without further RFI mitigation efforts, including observations within the FM and DTV bands. The results are compared to a previously published Low-Frequency Array (LOFAR) RFI survey. The remote location of the MWA results in a substantially cleaner RFI environment compared to LOFAR's radio environment, but adequate detection of RFI is still required before data can be analysed. We include specific recommendations designed to make the SKA more robust to RFI, including: the availability of sufficient computing power for RFI detection; accounting for RFI in the receiver design; a smooth band-pass response; and the capability of RFI detection at high time and frequency resolution (second and kHz-scale respectively).
\end{abstract}

\maketitle

\label{firstpage}


\section{Introduction}
Recent years have seen growth in the impact of radio-frequency interference (RFI) on radio astronomy, due to increased number of transmitters and wider bandwidths of radio observatories. Spectrum allocation management and radio-quiet zones help to limit the interference, but do not include all terrestrial transmissions, nor solve interference from air-born or satellite transmitters or accidental electromagnetic radiation, for example from cars or wind turbines.

While there has been some success in RFI mitigation by actual removal of the interference while retaining the underlying data \citep{spatial-filtering-parkes-multibeam-for-pulses,rfi-spatial-processing-hellbourg-2014}, such an approach is often not feasible due to technical limitations or the type of RFI. Therefore, a common approach is detection and flagging of contaminated data and ignoring these samples in further data analysis \citep{statistical-rfi-removal,pieflag-middelberg-2006,post-correlation-rfi-classification,prasad-flagcal-2012,serpent-peck-2013}. The consequences of this approach are that a certain fraction of data is lost due to interference, and frequency channels that are continuously occupied by transmitters cannot be observed. It is important to analyse the impact of RFI on a specific instrument to optimize the RFI mitigation approach. An increased understanding of the RFI situation will also help in several other ways: understanding the effects of RFI on the science; observation scheduling; designing robust hardware; choosing locations of future telescopes with a maximal cost/benefit approach; and designing effective spectrum management strategies.

In this article, we will look specifically at the RFI situation of the Murchison Widefield Array (MWA; \citealt{mwa-design-2009,mwa}). The MWA is a low-frequency array consisting of 128 tiles, each tile comprising of a 4x4 array of dual-polarization dipoles, which allow observing between 72--300~MHz with a 30.72-MHz instantaneous bandwidth; one of its main science drivers is to detect redshifted 21-cm radio signals from the Epoch of Reionization (EoR; \citealt{bowman-science-with-the-mwa-2013}). To avoid as much RFI as possible, the MWA is located at the CSIRO Murchison Radio-astronomy Observatory (MRO) in the Murchison region of Western Australia. Analysing the interference environment of the MWA may improve the MWA observing and processing strategy, and will additionally also provide valuable information for the SKA, because the cores of the SKA low-frequency aperture array are planned to be built in this vicinity.

Several levels of protection are in place to protect the radio quietness of the MRO. Within a 70~km radius, the Australian Communications and Media Authority\footnote{\url{http://www.acma.gov.au/}} and Western Australia government provide the strongest level of protection from other radio equipment across the frequency range 70~MHz to 25.25~GHz. Radio devices in this zone must not cause interference to radio astronomy.  Beyond 70~km, coordination zones extend out to~260 km radius at the lowest frequencies and reduce in size with increasing frequency. Individually licensed or spectrum-licensed radio devices in these zones must be coordinated with radio-astronomy requirements to eliminate or minimize interference. Situations not covered by the radio-quiet zone are: i) transmitters below 70~MHz; ii) most aircraft transmissions; iii) satellite transmissions; and iv) transmitters beyond 260~km from the centre of the MRO.

The interferometric arrays Low-Frequency Array (LOFAR; \citealt{lofar-2013}), Precision Array to Probe the Epoch of Reionization (PAPER; \citealt{parsons-paper-eorlimit-2014}) and the Giant Metrewave Radio Telescope (GMRT; \citealt{the-gmrt-swarup-chapter-2013}) observe in approximately the same frequency range as the MWA. For each of these instruments, projects are ongoing to detect redshifted 21-cm signals from the Epoch of Reionization. The overlap in frequency gives an opportunity to compare the observatory sites and hardware designs with respect to the RFI impact. So far, initial observations with the MWA have produced scientific results without RFI issues \citep{hurley-walker-mwacs-2014, mckinley-fornaxa-2014, hindson-cluster-emission-2014}, which of course is not surprising given its remote location. \citet{parsons-paper-eorlimit-2014} have shown that the radio environment for PAPER in the Karoo desert of South Africa is sufficiently clean to reach with long integration a $(41 \textrm{mK})^2$ upper limit for the EoR brightness temperature at one scale and redshift, two orders of magnitude (in $\textrm{mK}^2$) away from the expected EoR signal strength. \citet{lofar-radio-environment} show that for regular observations the radio environment of LOFAR does not pose unsurmountable issues, even though LOFAR is located in a populated area. This is confirmed by \citet{ncp-eor-yatawatta}, where the authors reach near-thermal noise sensitivity in the first EoR long-integration images with LOFAR. They argue that RFI is not a limitation for further increasing the sensitivity, and \citet{offringa-rfi-distributions} conclude that with sufficient precautions, such as good receiver design, accurate detection methods, and high time and frequency resolutions, residual RFI is weak and averages down in a similar way to Gaussian noise. However, none of the EoR projects have yet processed enough data to reach the sensitivities required for a detection of EoR signals, and low-level RFI could potentially prevent such a detection.

Recently, experiments to use the Moon as a calibrator for EoR experiments have shown that the reflection of terrestrial transmitters by the Moon complicates such an experiment \citep{mckinley-moon-2013}. However, reflections are spatially restricted to the centre of the Lunar disk, and the high resolution of LOFAR allows the separation of the reflected and intrinsic power from the Moon \citep{vedantham-2014-todo}. Another study shows that objects in space of about a meter in size, such as satellites and space debris, may also reflect terrestrial transmissions with enough strength to be observable by the MWA \citep{tingay-space-debris-2013}. While tracking space debris is a useful asset, such reflected RFI can be a problem for EoR experiments, especially all-sky experiments that try to measure the global EoR signal \citep{vedantham-2014-todo}.

In this paper, we will describe the mitigation strategy implemented for the MWA, and show examples of RFI found and results of the RFI detection. Sect.~\ref{ch:method} describes the approach taken for the MWA, including software, algorithms and computational challenges involved. This strategy is applied to 63~h of MWA data. In Sect.~\ref{ch:data-description}, these data are described. Sect.~\ref{ch:detection-results} presents examples of RFI that were found in these data, as well as the efficacy of the RFI detection. In Sect.~\ref{ch:lofar-comparison}, the results are compared to a previously performed LOFAR RFI survey. Finally, in Sect.~\ref{ch:conclusions-and-discussion}, conclusions are drawn and discussed.

\section{Method} \label{ch:method}
RFI detection, often referred to as ``data flagging'', is one of the first steps in processing the data from any interferometer. One measure of the performance of an RFI detection method is its accuracy, which is often quantified by the average number of false-positive and true-positive detections resulting from the method. It is important to perform initial RFI detection and excision at high time and frequency resolution, because this increases detection accuracy and decreases the loss of data \citep{lofar-radio-environment}. Consequently, RFI detection has to work on large data volumes, and its computational cost is therefore a concern --- in particular for many-element arrays. In some projects, simple amplitude thresholding is used to mitigate the worst interference, which is computationally cheap but not very accurate. For example, a $3\sigma$ threshold is used for analysing PAPER data in \citet{parsons-paper-eorlimit-2014}. Several observatories or projects have designed pipelines that include more advanced RFI mitigation. Examples of such pipelines include \textsc{aoflagger} \citep{post-correlation-rfi-classification,scale-invariant-rank-operator}, originally designed for LOFAR; \textsc{flagcal} for preprocessing data from the Giant Metrewave Radio Telescope (GMRT; \citealt{prasad-flagcal-2012}); \textsc{pieflag} \citep{pieflag-middelberg-2006} and \textsc{mirflag} \citep{lenc-mirflag-2010} mostly used for the Australia Telescope Compact Array (ATCA); and \textsc{serpent} for preprocessing data from the Multi-Element Radio Linked Interferometer Network (e-MERLIN; \citealt{serpent-peck-2013}). For RFI detection in MWA observations, LOFAR's \textsc{aoflagger} is used. It has been shown that this flagger has a good accuracy and is fast \citep{lofar-radio-environment}. It also has a library interface\footnote{The documentation for the \textsc{aoflagger} library interface can be found at \url{http://aoflagger.sourceforge.net/doc/api/}}, which allows it to be integrated in a pipeline.

\subsection{The \textsc{aoflagger} RFI detector}

\begin{figure}
\begin{center}\hspace*{-0.8cm}\includegraphics[width=10.5cm]{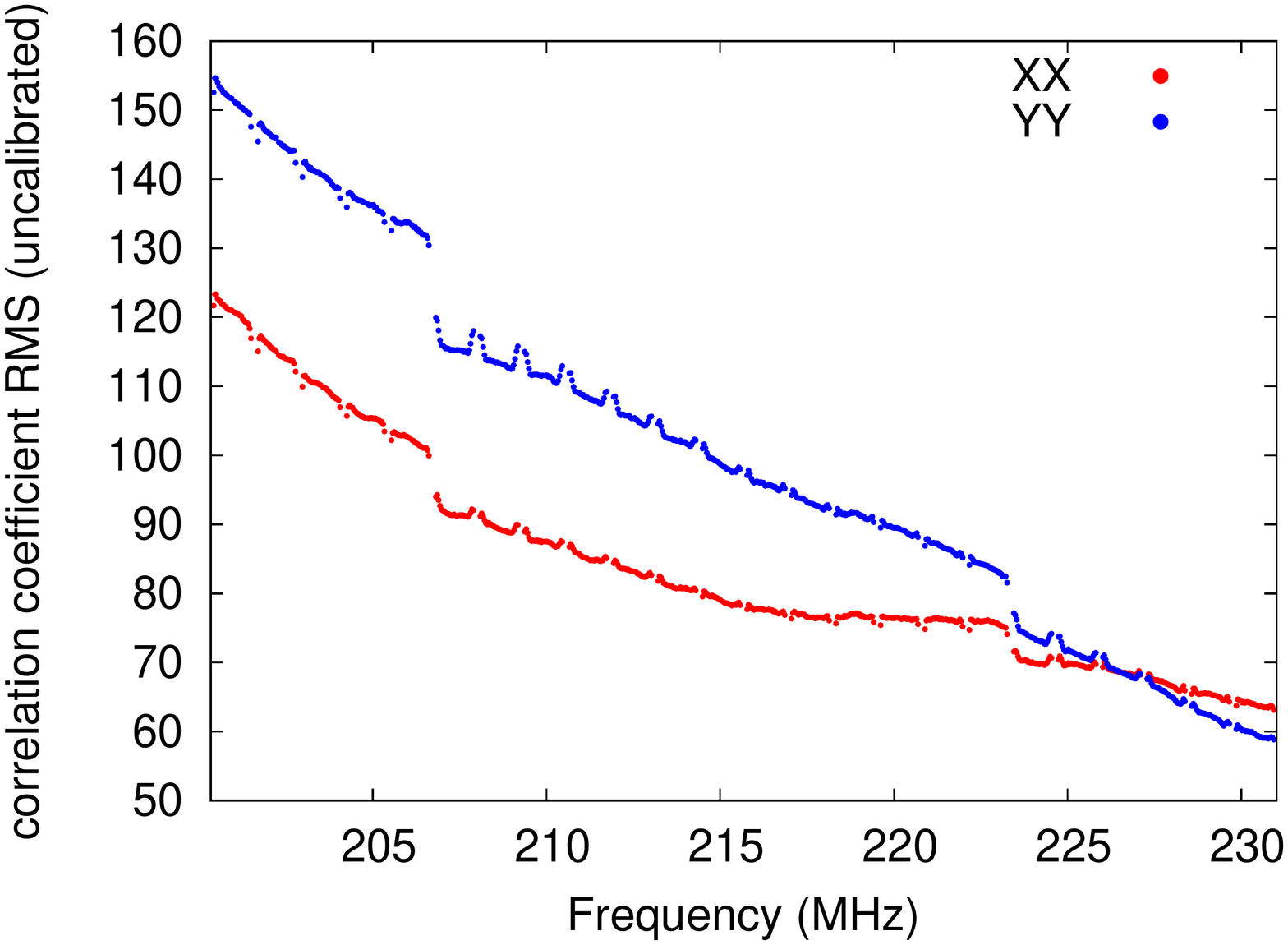}
\caption{Correlator output RMS with respect to frequency in an MWA high-frequency observation, calculated over all cross-correlated baselines and 112 seconds of data. In this band, the band-pass shows two large discontinuities over frequency, because the MWA receivers apply different digital gains at different frequencies to minimize quantization noise. The 1.28~MHz sub-bands have already been corrected for the band-pass shape of the poly-phase filter, but a residual 1.28-MHz pattern is visible due to aliasing.
}
\label{fig:bandpass}
\end{center}
\end{figure}

To detect RFI in MWA observations, we have used \textsc{aoflagger} and implemented this as a standard MWA tool to flag all MWA data. \textsc{aoflagger} is a general-purpose RFI flagging tool developed originally for LOFAR \citep{lofar-radio-environment}. Specific customizations, such as changing the threshold levels and expected smoothness of good data, can be made for different telescopes to optimize the detection accuracy for different band-pass shapes, time and frequency resolutions, and fields of view. Strategies for several telescopes have been implemented in the \textsc{aoflagger} software, including an MWA-specific strategy which is used in this work. In the LOFAR strategy, the sky contribution is estimated by applying a 2-dimensional high-pass filter on the visibility amplitudes of each baseline in the time and frequency domains. Subsequently, line-shaped features are detected by the SumThreshold method, which is a combinatorial threshold method \citep{post-correlation-rfi-classification}. After iterating these steps a few times, the scale-invariant rank (SIR) operator is applied on the two-dimensional flag mask. The SIR operator is a morphological technique to search for contaminated samples \citep{scale-invariant-rank-operator}.

The MWA and LOFAR strategies differ qualitatively in one aspect. For the MWA, an extra bandpass correction is added, which is performed by dividing the bandwidth into 48 equal sub-bands and dividing each sub-band by its Winsorized standard deviation (see \citet{variance-estimates} for an explanation of Winsorized statistics). This step is required to smooth out discontinuities due to varying digital gains over frequency that are applied by the receivers to minimize quantization noise. An example of these discontinuities is shown in Fig.~\ref{fig:bandpass}. The bandpass corrections are not permanently applied to the data, but only used during flagging. Normally, a per-channel calibration is performed after flagging in order to obtain accurate flux density calibration, which corrects the gain discontinuities permanently. Recently, the applied digital gains have been smoothed to prevent these discontinuities, which makes it possible to skip the band-pass correction step, but it was necessary to apply this correction for the data used here.

In this work, \textsc{aoflagger} version 2.6 released on 26 June 2014 is used.

\subsection{\textsc{cotter}: the MWA preprocessing pipeline}

The \textsc{aoflagger} software provides a C++ library that can be integrated in a pipeline such that intermediate data can be kept in memory. This minimizes the reading and writing of data. We have written an MWA-specific preprocessing pipeline named \textsc{cotter}\footnote{Etymology: Cotter is a geographical area around Cotter River, near Mount Stromlo Observatory in Canberra.} that uses the \textsc{aoflagger} library for RFI detection. RFI detection is only performed on cross-correlations. Auto-correlations are normally ignored, because they are not used in imaging. In addition to the RFI detection, \textsc{cotter} performs the following steps: it converts the raw correlator files into \textsc{casa} measurement sets or \textsc{uv-fits} files; applies bandpass gain corrections; corrects the phases for varying cable lengths; calculates the $u,v,w$-coordinates; applies phase tracking to the desired sky coordinates; flags samples from the correlator that are missing or incorrect; and allows averaging the visibilities in frequency and/or time to reduce the data volume. It also collects various statistics and writes these into a measurement set using the LOFAR quality statistics format\footnote{Described in ``MeasurementSet description for LOFAR version 2.08'' by Schoenmakers \& Renting.}. Tools are available to analyse these statistics, e.g. the \textsc{aoqplot} tool that is part of the \textsc{aoflagger} software can plot the statistics over various dimensions in an interactive manner.

MWA observations are split into snapshots of a few minutes by the correlator. The MWA archive stores the raw correlator outputs for each snapshot as an observation that can be referenced by its observation ID. For more details about the correlator, see \citet{ord-2014-mwa-fpga-gpu}. Currently, the \textsc{cotter} preprocessing pipeline is run by the scientist that calibrates and images the data. Once the scientist has downloaded the raw files for a given observation ID, there are various ways of processing MWA data. For imaging MWA data with the Real-Time System (RTS; \citealt{rts-mwa-2008}), \textsc{cotter} is run in a special mode such that it only flags the data, and does not convert the raw correlator files. After running \textsc{cotter}, the raw files and a flag mask are given as input to the RTS. For data processing with the Common Astronomy Software Applications (\textsc{casa}; \citealt{casa-2008}), \textsc{miriad} \citep{miriad-sault-1995}, \textsc{wsclean} \citep{offringa-wsclean-2014} and/or the Fast Holographic Deconvolution software (\textsc{fhd}, \citealt{fhd-sullivan-2012}), the \textsc{cotter} output is set to either the \textsc{casa} or \textsc{uv-fits} format. The output is then directly readable by the most common astronomical software packages. 

One particular issue in implementing \textsc{cotter} is that both the raw correlator files and the desired output files are ordered in time, but flagging is done baseline by baseline. The SumThreshold and SIR operator algorithms that are used by the flagging strategy use statistics calculated over large time and frequency ranges. Therefore, detection accuracy is improved when the time-frequency flagging intervals are as large as possible. However, a typical snapshot is about 50 GB in size and without increasing disk I/O overhead it requires 50 GB of memory to perform flagging on the full data. When less memory is available, \textsc{cotter} will split the observation into a number of shorter time segments and flag these independently. This is similar to the partitioning that is used in the \textsc{ndppp} software used by LOFAR \citep{lofar-imaging-cookbook} to overcome the time-ordering problem. Partitioning the data has the undesirable consequence that executing \textsc{cotter} on a low-memory machine decreases its flagging accuracy. As not all astronomers have easy access to large-memory machines, a platform with sufficient memory has been set up that runs a partial \textsc{cotter} preprocessing on all observations. In this use-case, \textsc{cotter} is run in a flagging-only mode on the large-memory machine. The astronomer downloads the raw files and the flagging output files, and reruns \textsc{cotter} to apply the RFI detection from the first run and convert the raw files to his/her preferred output format.

When time or frequency averaging is requested, \textsc{cotter} averages samples together that have not been flagged. When all input samples are flagged, the average of all input visibilities is stored into the output sample, and the output sample is flagged. This method makes it possible to superficially analyse flagged samples in the output, even though information is lost in the averaging. \textsc{cotter} stores a weight for each visibility in the output file, which is scaled to the number of unflagged input samples that were used for the output sample. Because of this, when no averaging is requested, the output is 50\% larger than the input (one extra float per complex float visibility). In practice, most MWA observations are recorded at a resolution of 0.5~s and 40~kHz, and averaged to a resolution of 2--4 s and 40--80 kHz to reduce the data volume. This does not cause any significant time or bandwidth decorrelation up to the first beam null for MWA data. \textsc{cotter} performs the phase shifting and cable delay corrections before averaging, and recalculates the $u,v,w$-values for the central time and frequency of the output sample. This helps to prevent time/frequency decorrelation. The central time/frequency of an output sample is set to the time/frequency mean of the corresponding input samples, independently of what input samples are flagged.

\begin{table*}
\caption{List of observations used in the analyses. The RFI column contains the fraction of visibilities that the initial \textsc{aoflagger} run has classified as RFI (including false positives, excluding data loss caused by band edges). Occasionally, interference from digital TV (DTV) is visible. The DTV column displays the fraction of visibilities that were unusable because of the presence of DTV signals, and which the flagger does not flag. Not all observations cover the DTV frequencies.}\label{tbl:obs-list}
\begin{center}%
\begin{tabular}{|l|l|r|r|r|c|}
\hline
\textbf{Project} &\textbf{Date} & \textbf{Frequencies (MHz)} & \textbf{Duration} & RFI & DTV \\
\hline
EoR high& 2013-08-23 & 167.0--197.7              & 3~h & 0.53\% & 0\% \\
GLEAM & 2013-08-25 & 72.3--133.8, 138.9--230.8 & 7~h & 0.94\% & 0\% \\
EoR low& 2013-08-26 & 138.9--169.6              & 6~h & 0.81\% & ---\\
GLEAM & 2013-11-05 & 72.3--133.8, 138.9--230.8 & 7~h & 1.27\% & 0\% \\
GLEAM & 2013-11-25 & 72.3--133.8, 138.9--230.8 & 7~h & 0.69\% & 0\% \\
EoR high& 2014-02-05 & 167.0--197.7              & 6~h & 0.54\% & 0\%\\
GLEAM & 2014-03-16 & 72.3--133.8, 138.9--230.8 & 7~h & 0.79\% & 0\% \\
GLEAM & 2014-03-17 & 72.3--133.8, 138.9--230.8 & 7~h & 1.64\% & 1.29\% \\
EoR high& 2014-04-10 & 167.0--197.7              & 6~h & 0.68\% & 0\%\\
GLEAM & 2014-06-18 & 72.3--133.8, 138.9--230.8 & 7~h & 0.98\% & 0\% \\
Orbcomm RFI test&2014-08-27& 131.2--161.9     & 2$\times$8~m& 1.85\% & --- \\
\hline
\multicolumn{3}{|r|}{\textbf{Total}} & \textbf{63h} & \textbf{0.96\%} & \textbf{0.14\%}\\
\hline
\multicolumn{4}{|r|}{\textbf{With uniform channel coverage}} & \textbf{1.13\%} & \\
\hline
\end{tabular}
\end{center}
\end{table*}

\section{Data description} \label{ch:data-description}
We analyse the RFI in 10 nights of data from two different MWA projects: the MWA EoR project \citep{bowman-science-with-the-mwa-2013} and the GaLactic and ExtrAgalactic MWA (GLEAM) survey (Wayth et al., in prep.). The schedule of night-time observations is listed in Table~\ref{tbl:obs-list}.

The MWA EoR project observes primarily in the 138.9--197.7~MHz range, covering the HI 21-cm line with redshift 6.1--9.2. The MWA has a 30.72-MHz instantaneous bandwidth, which makes it necessary to observe at two different frequencies to cover the desired EoR bandwidth. An observing night is centred on either 154.2 or 182.4~MHz, which results in an overlap of 2.6~MHz. The bands covered by these central frequencies will be referred to as the EoR low and high bands, respectively. For the EoR observations, the beam formers are changed every 30~minutes to track the selected field with the primary tied-array beam, such that the sensitivity towards the field is maximal. The GLEAM survey covers 72.3--230.8~MHz split into 5 bands of 30.72~MHz. The GLEAM observations are made in drift-scan mode, with the 5 bands rotated through in sequence on a 2-minute cadence.

Both projects avoid the sub-bands that cover the frequency range 133.8--138.9~MHz, because the ORBCOMM low-Earth-orbiting (LEO) satellites transmit in these sub-bands. When these frequencies are observed, the sub-bands are often so strongly contaminated that it affects imaging sensitivity. For completeness, we include a 16-min observation that covers this frequency range. Our analyses do not include frequencies above 231~MHz, although observations above 231~MHz are possible with the MWA. Data above 230~MHz have shown contamination from the constellation of Milstar communication satellites.

The poly-phase filter bank that performs the first separation into sub-bands introduces a 1.28~MHz periodic spectral signature. As shown in Fig.~\ref{fig:bandpass}, the poly-phase filter band-pass shape is hard to remove completely, because of aliasing each sub-band is affected by leakage from signals in its adjacent sub-bands. The leakage from adjacent sub-bands manifests itself as if the sub-band band-pass is direction dependent. To solve this, the bordering 80~kHz on both sides of a 1.28~MHz subband are normally flagged by \textsc{cotter}. This implies that in normal observations, 13\% of the data are lost due to the poly-phase filter. However, while 80~kHz is sufficient to prevent imaging artefacts, we noticed that the RFI statistics were still slightly biased by the edge channels. In particular, the detected fraction of RFI over frequency is about 0.5\% higher (i.e., $\sim1.5\%$ instead of $\sim1\%$) in the edge channels after flagging 80~kHz of the edges. Therefore, for the analyses in this paper we increase the removed bandwidth to 200~kHz on either side of each 1.28~MHz sub-band, or 32\% in total.
\noindent\begin{figure*}
\begin{center}\hspace*{-0.2cm}\includegraphics[width=18cm]{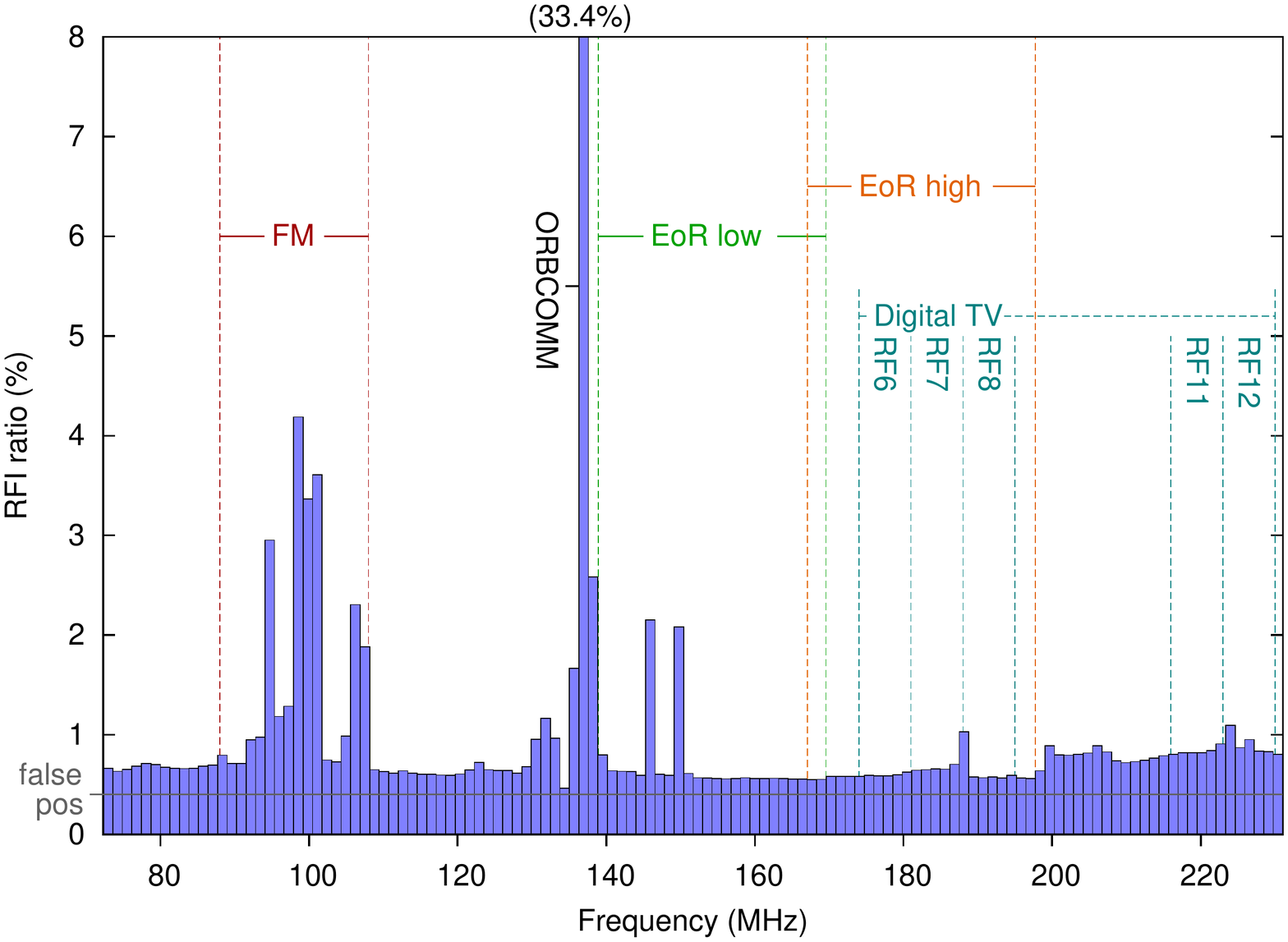}
\caption{RFI occupancy per subband, calculated over all observation nights except GLEAM 2014-03-17. The latter has been left out because it is affected by DTV. The horizontal gray line represents the false-positives rate of the RFI detection. The RFI fractions are consistently higher than the false-positives rate because of transient broad-band RFI.}
\label{fig:rfi-per-band}
\end{center}
\end{figure*}

\noindent\begin{figure*}
\begin{center}\hspace*{-0.2cm}\includegraphics[width=18cm]{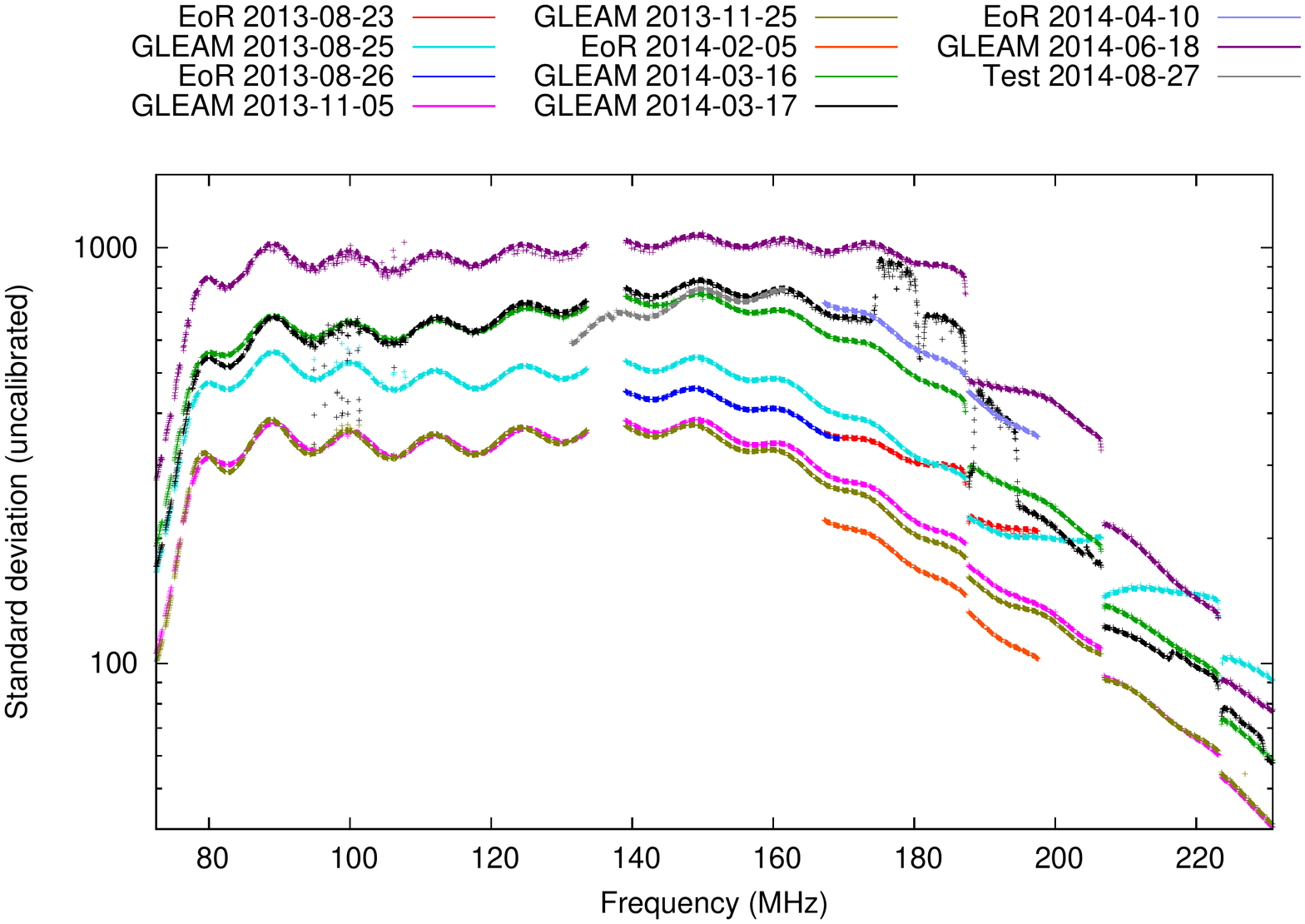}\vspace{-1cm}
\caption{Observed visibility RMS in the high-resolution (40~kHz) channels for each observation. This is the residual noise in the visibilities after RFI excision. A few channels in the FM bands show an abnormal standard deviation, and observation `GLEAM 2014-03-17' shows DTV contamination around 180~MHz. The variability over nights is caused by the different celestial observing times and pointing directions, and therefore the difference in apparent brightness of e.g. the Galaxy.}
\label{fig:stddev-per-set}
\end{center}
\end{figure*}

\section{Detection results} \label{ch:detection-results}
The ``RFI'' column of Table~\ref{tbl:obs-list} lists the fraction of samples that were classified as RFI by the \textsc{aoflagger} in the cross-correlations for each observing night. This includes false detections, which are estimated in \citet{lofar-radio-environment} to account for approximately 0.4\% of detections. Compared to the EoR observations, the GLEAM observations have a higher average of 1.05\% of RFI. This is mostly caused by the FM bands, which are only observed in GLEAM observations. The EoR low-band night shows 0.81\% RFI, while the EoR high-band observations have an average of 0.58\%. The total RFI occupancy in all observations is 0.96\%. Weighting the occupancy in each channel by the inverse time that it is observed results in a global RFI occupancy of 1.13\% in the 72.3--230.8 MHz range.

Fig.~\ref{fig:rfi-per-band} shows the overall detected RFI occupancy per sub-band, as calculated over all observations except GLEAM 2014-03-17. The latter is the only night affected by interference from digital TV (DTV), and will be analysed later in this section. RFI occupancy is calculated as the percentage of discrete visibilities that are detected as RFI by the flagger at the resolution of the correlator output. The FM bands around 100~MHz and the ORBCOMM bands around 138~MHz are clearly present in the data. Excluding the RFI from DTV, the EoR high band is slightly cleaner than the EoR low band, and its worst subband at 188.2 MHz has 1.03\% occupancy. The sub-bands at 145.9 and 149.8~MHz in the EoR low band have both 2.1\% occupancy.

The residual noise levels after flagging can be used to validate whether the flagged data are free of RFI. In Fig.~\ref{fig:stddev-per-set}, the residual noise levels are plotted per high-resolution channel for each of the observations. Observation `GLEAM 2014-03-17' shows residual DTV interference, both in the frequency range 174--195~MHz (radio frequencies (RF) 6, 7 and 8) and 216--230~MHz (RF 11 and 12), and it is clear that this RFI has not been adequately flagged. Therefore, DTV interference has to be detected with another method. Additionally, some channels in the FM radio band show higher standard deviations as a result of the smaller amount of available data after flagging and possibly because of RFI leakage. Nevertheless, because the effect is small these frequencies can be calibrated and imaged without a problem. FM-band RFI is noticeably worse when pointing at the southern horizon, however beyond this we do not have sufficient data to explore any correlation between pointing direction and RFI. The subband at 137~MHz that is occupied by ORBCOMM is hard to calibrate because of the small amount of residual data per channel, and possibly also because of residual RFI. With the exception of the ORBCOMM frequencies and DTV affected nights, the RFI detection employed in \textsc{cotter} is sufficient to allow calibration and imaging without further RFI mitigation efforts. This has been verified by early imaging results from the GLEAM survey and the MWA commissioning survey \citep{hindson-cluster-emission-2014, hurley-walker-mwacs-2014, murphy-mwa-exoplanets-2014}.

\noindent\begin{figure*}%
\begin{center}\hspace*{-0.2cm}\includegraphics[width=10cm]{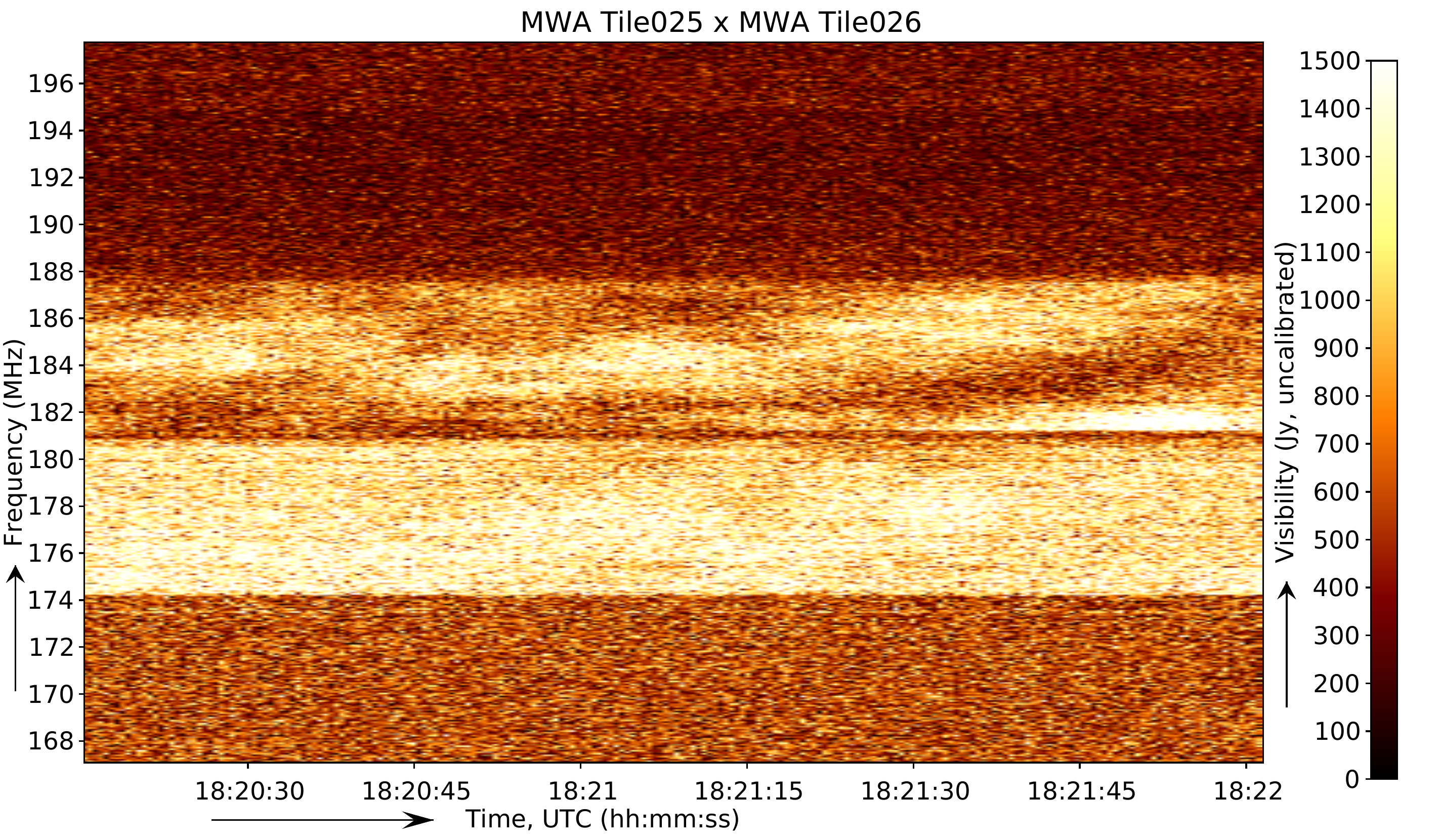}\includegraphics[width=5cm]{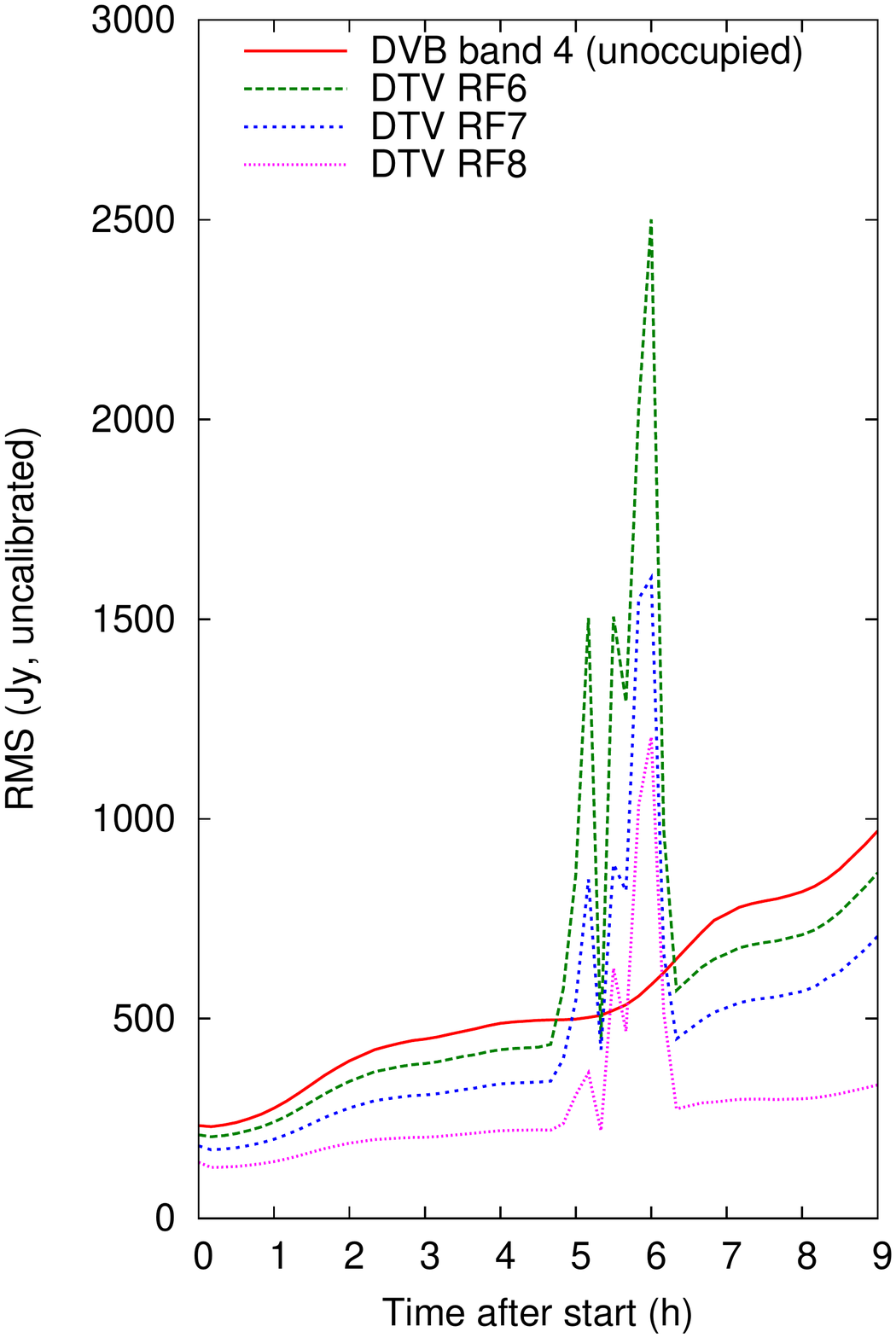}
\end{center}
\caption{RFI from digital TV (DTV), which is visible in 1 of the 9 observations (GLEAM~2014-03-17) that cover the DTV frequencies. Left plot: dynamic spectrum for the worst affected snapshot; right plot: visibility RMS in a few of the DTV bands for the affected night after flagging. Because of its broadband nature, this kind of RFI is not well-flagged in the initial \textsc{aoflagger} step, but it is detectable in the global statistics. }
\label{fig:dvb}
\end{figure*}

Because significant DTV interference residuals are visible in `GLEAM 2014-03-17' after flagging, we have analysed this night more extensively. The DTV transmitters are terrestrial, and the fact that only this night observes the DTV, implies the RFI must originate from an over-the-horizon transmitter that is reflected by unusually strong ionospheric activity or tropospheric ducting. As can be seen in the left plot of Fig.~\ref{fig:dvb}, this kind of RFI can fully contaminate frequencies 174--195~MHz, thus over half of the instantaneous bandwidth. Because the \textsc{aoflagger} determines its thresholding levels from the data, and because it needs to be insensitive to steep RMS jumps over frequency due to the varying coarse channel gains (see Fig.~\ref{fig:bandpass}), it is insensitive to RFI covering such a broad spectrum.

Although the flagger does not adequately flag DTV interference, the right plot of Fig.~\ref{fig:dvb} shows that from the data statistics it is possible to determine whether a snapshot is affected by DTV interference. An option is to test whether the visibility RMS of a snapshot in any of the occupied DTV bands exceeds the RMS in the Digital Video Broadcasting (DVB) band 4 (167--174~MHz), which is not used for broadcasting. These statistics are calculated by \textsc{cotter}, hence this can be validated without having to read the data again. When DTV is detected in a snapshot, the entire snapshot can be removed or the residual unaffected data within the snapshot can be used. This could be dealt with by accounting for lower SNR and a decreased $uv$ coverage, however putting such a mechanism in place (rather than just deleting affected snapshots entirely) may not be worth the effort. Because 2 out of 7 hours are affected in 1 out of the 9 randomly selected nights, the probability that DTV interference occurs is roughly 3\%. The bands involved are RF 5, 6, 7, 11 and 12, each of 7~MHz bandwidth. Therefore, when DTV interference occurs, it affects 35 of the 159~MHz bandwidth. The visibility ratios that are lost because of DTV RFI are detailed in the ``DTV'' column of Table~\ref{tbl:obs-list}. Because these frequencies overlap with the 21-cm HI frequencies redshifted to the EoR, in reality these frequencies are observed more often, so the actual loss is somewhat higher.

\noindent\begin{figure*}
\begin{center}\hspace*{-0.2cm}\includegraphics[width=18cm]{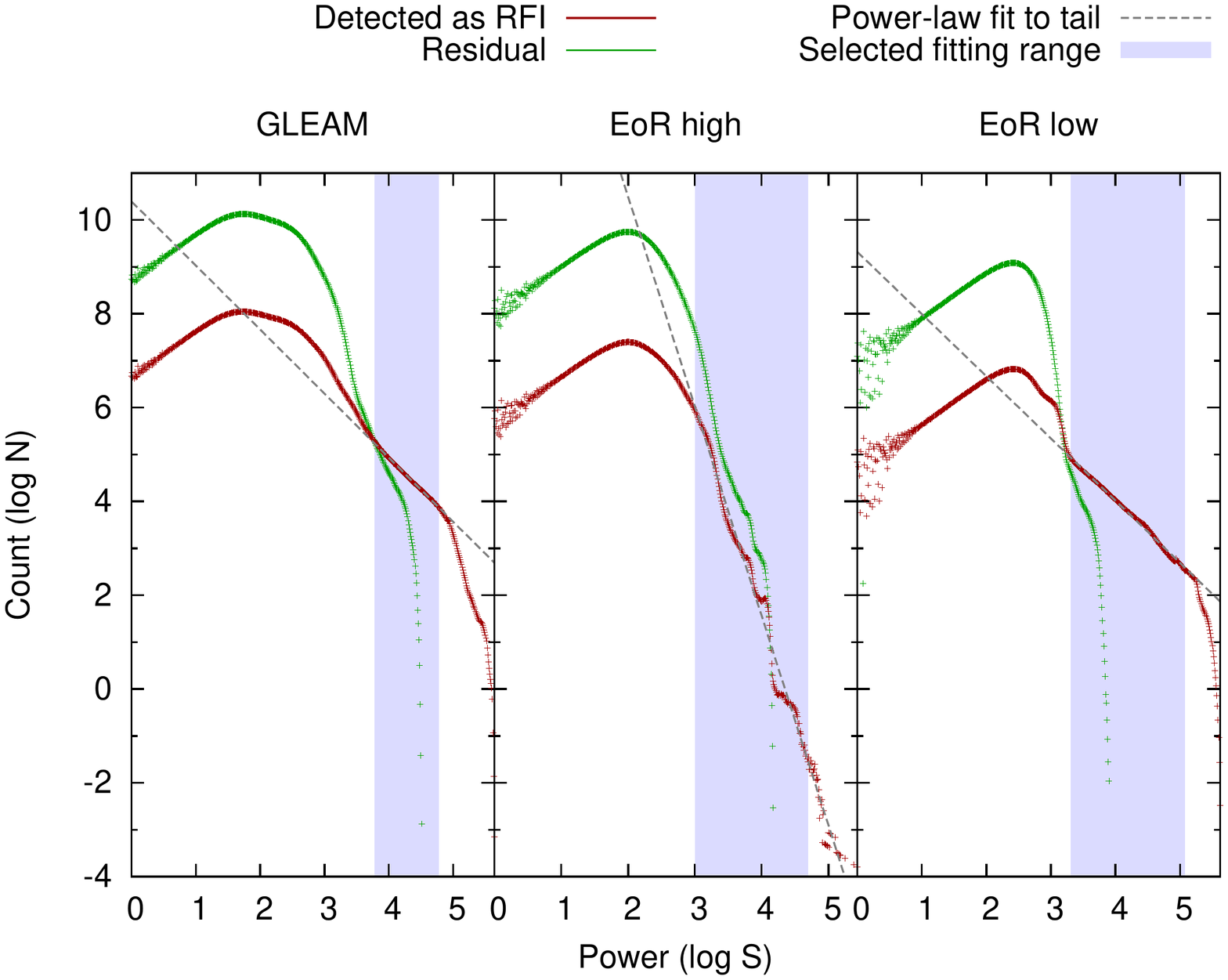}
\caption{Visibility amplitude distributions with logarithmic axes. The dashed line represent fits to the function $N=\beta S^\alpha$ over a reasonably constant part of the tail (gray area, selected by eye) of the RFI distribution, resulting in $\alpha=-1.37$ for GLEAM, $\alpha=-4.47$ for EoR high, and $\alpha=-1.33$ for EoR low.}
\label{fig:lognlogs}
\end{center}
\end{figure*}

\subsection{Distribution analysis}
Using LOFAR, \citet{offringa-rfi-distributions} show that when a uniform spatial distribution with sufficient interfering transmitters is observed, the distribution of the visibility brightness will have a power-law tail described by $N\propto S^\alpha$, where $N$ is the differential number of visibilities, $S$ is the visibility brightness and the exponent $\alpha$ is found by \citet{offringa-rfi-distributions} to be typically -1.5 to -1.6. For the GLEAM, EoR high-band and EoR low-band observations studied here, the distributions are shown in Fig.~\ref{fig:lognlogs}. The three distributions are different due to the fact that observations cover different frequencies.  Separate distributions are shown for visibilities that have been detected as RFI (red lines) and those that have been classified as RFI-free (green lines). The EoR high-band curves do not include the night that was contaminated by DTV RFI.

The distribution curves of RFI-detected samples do not show a very significant power-law tail. Fits over regions selected by eye have been overlaid in Fig.~\ref{fig:lognlogs}. The EoR high-band observes too little RFI to generate such a power-law tail, which is reflected by its fit of $N\propto S^{-4.47}$. The GLEAM and EoR low-band observations do show a small power-law component but the fitted exponent depends strongly on what part of the tail is selected. The selected regions result in $N\propto S^{-1.37}$ and $N\propto S^{-1.33}$ for the GLEAM and EoR-low curves, respectively. Due to the small region over which the power laws hold, as well as the possible high error in the fits because of the subjective data selection, it is hard to infer if the transmitters have a uniform spatial distribution.

In the ideal case, the residual visibility distribution would show a smooth Rayleigh curve (see \citealt{offringa-rfi-distributions}). The residual curves in Fig.~\ref{fig:lognlogs} do not fall off with RFI power as quickly as a Rayleigh curve would, and therefore the distributions have a slight excess of samples with higher amplitudes. This excess is caused by the variable nature of the data, for example because the noise level increases when the Galaxy goes through the beam. The EoR-high distribution has some extra features in its tails that are not smooth. Further analyses showed that these are also caused by the Galaxy, causing a few short baselines to observe samples with high amplitudes. Visual inspection of snapshots with a non-smooth residual distribution tail did not show residual RFI in such sets.

\noindent\begin{figure*}
\subcaptionbox{RFI contamination found in the 2 m amateur band (146 MHz).}[\linewidth]{%
\includegraphics[width=9.1cm]{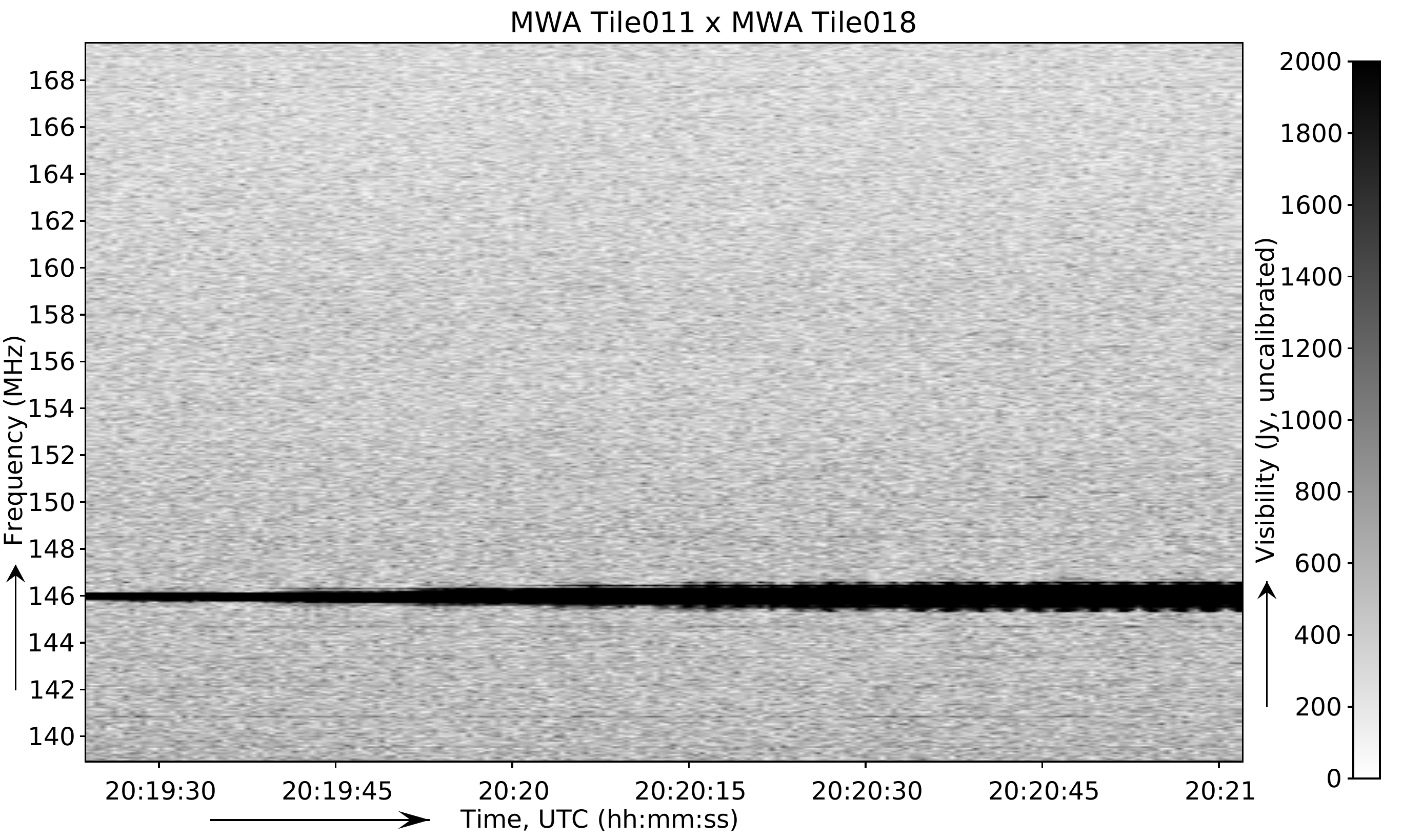}\includegraphics[width=9.1cm]{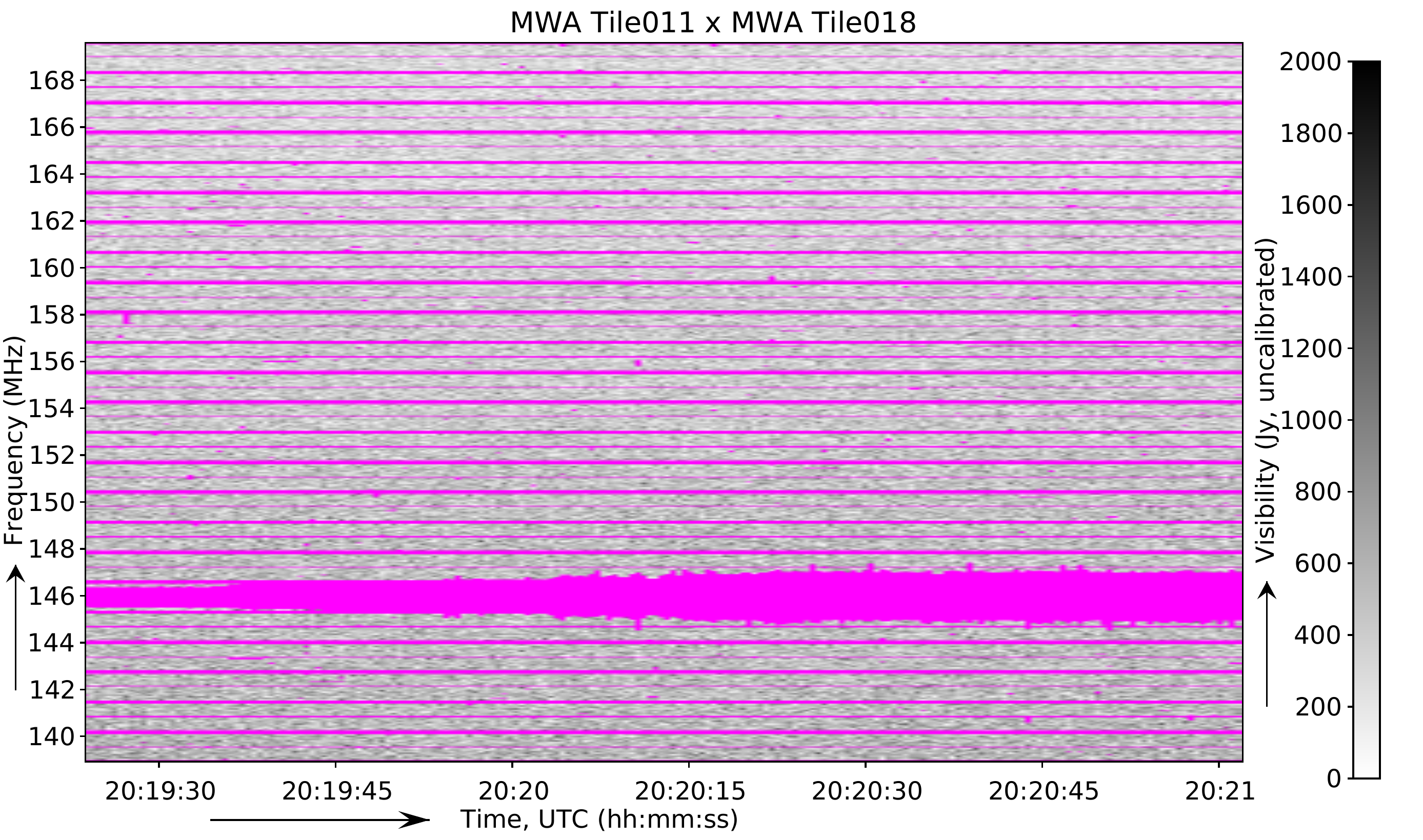}
\label{fig:amateur-2m}
}\\\vspace{3mm}%
\subcaptionbox{Short RFI burst of 2 s centred on 150.17 MHz. Detection of this kind of RFI requires flagging at high time and frequency resolution.}[\linewidth]{%
\includegraphics[width=9.1cm]{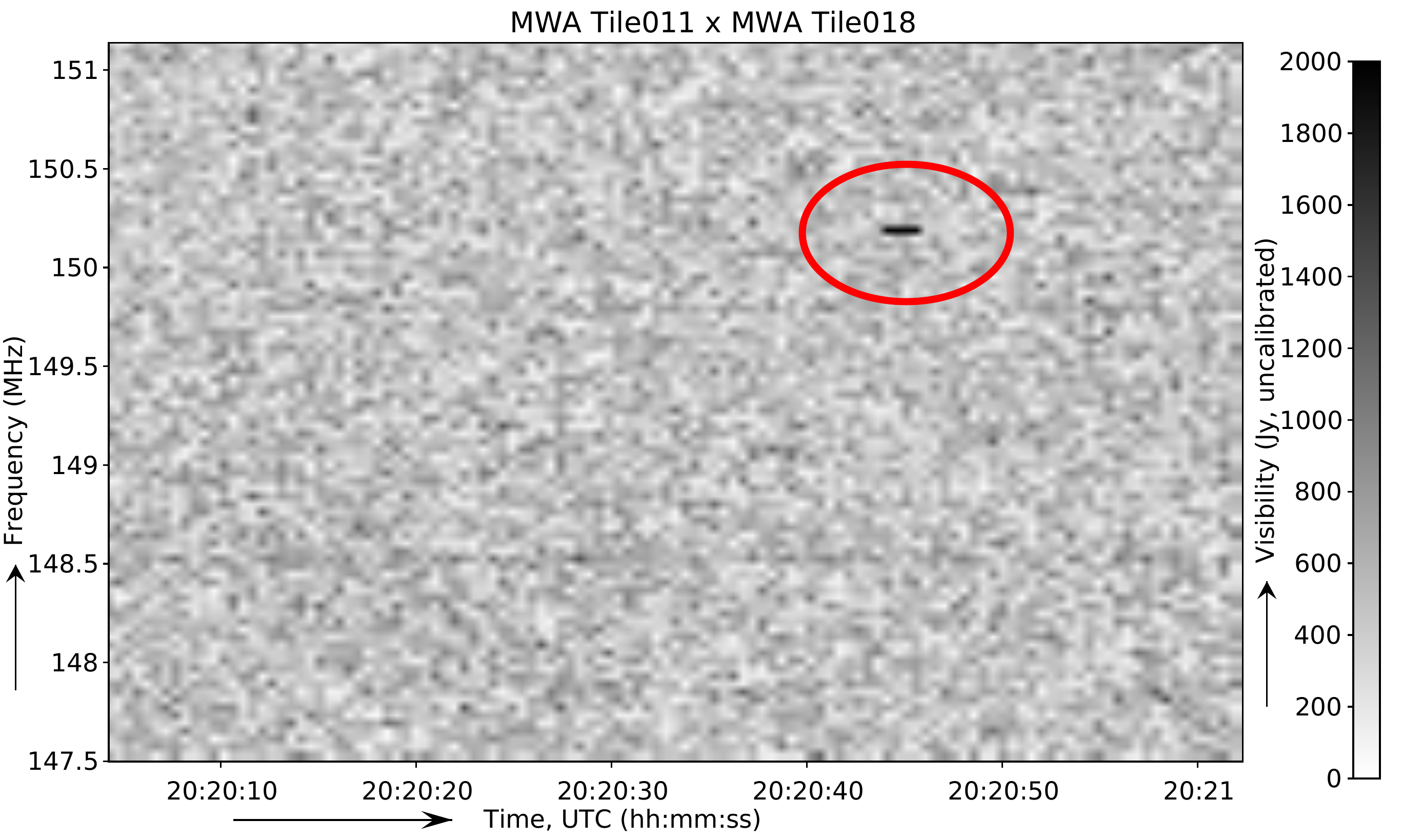}\includegraphics[width=9.1cm]{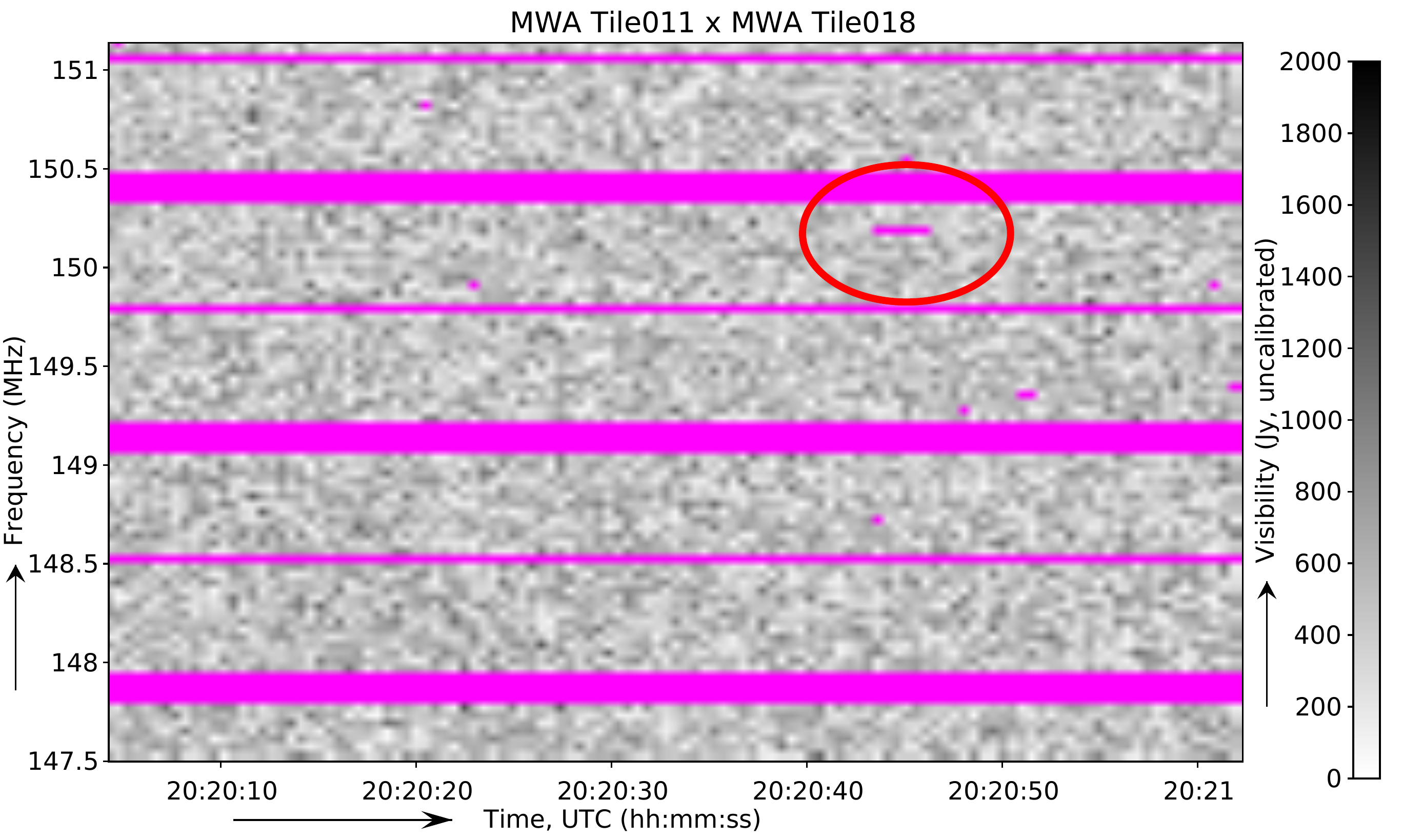}
\label{fig:150_2}
}%
\caption{RFI events found in the EoR low band (138.9--169.6 MHz) in a 2-min snapshot with relatively high RFI contamination. These panels show the Stokes I amplitudes. In the right figures, the result of RFI detection is shown with purple. The horizontal flagged lines are flagged because they are 1.28-MHz subband edge or centre channels, which are unusable because of aliasing of the poly-phase filter bank and DC offsets, respectively.}
\label{fig:2m-amateur-band}
\end{figure*}

\noindent\begin{figure*}%
\begin{center}\hspace*{-0.2cm}\includegraphics[width=8cm]{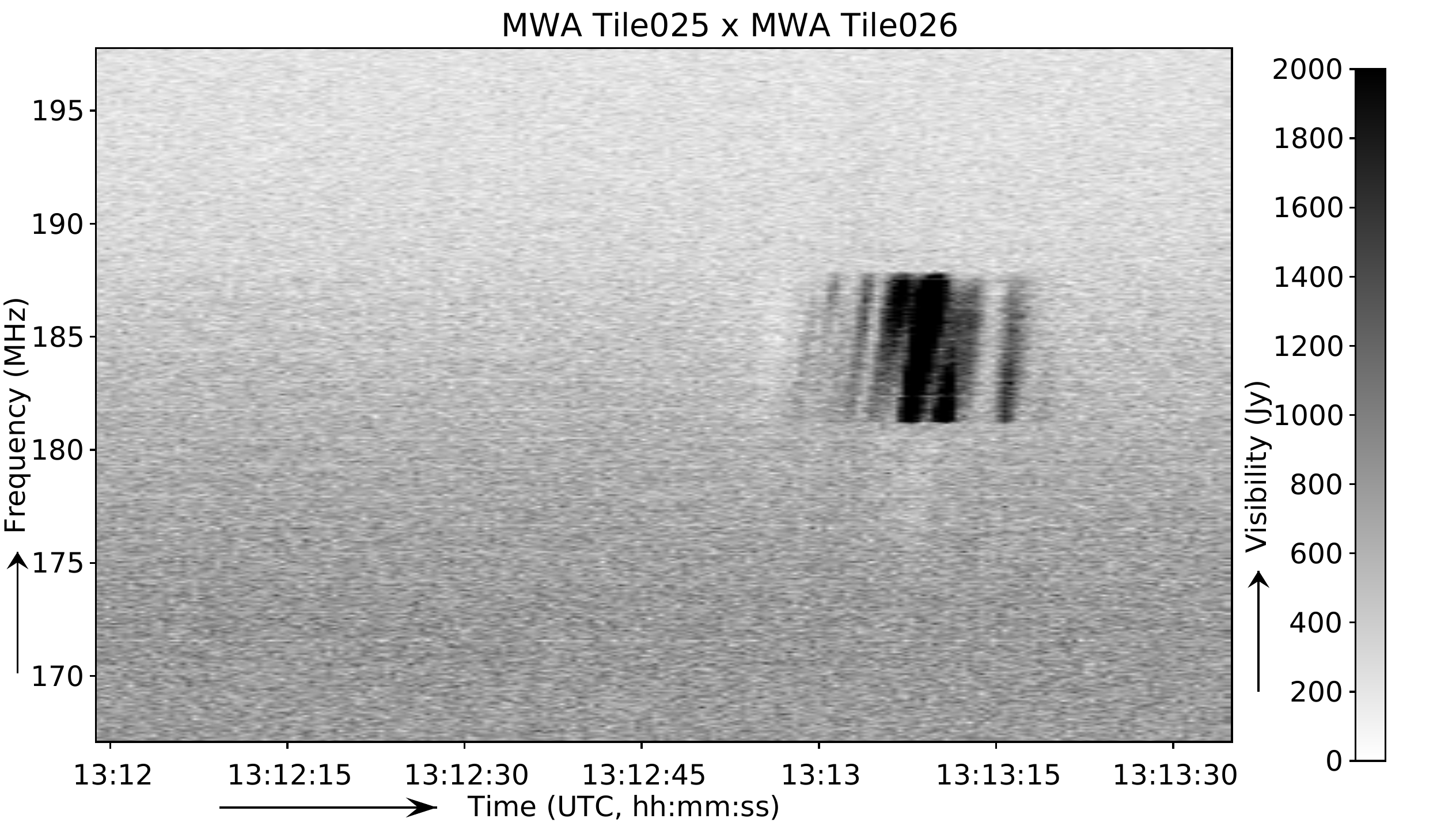}\includegraphics[width=8cm]{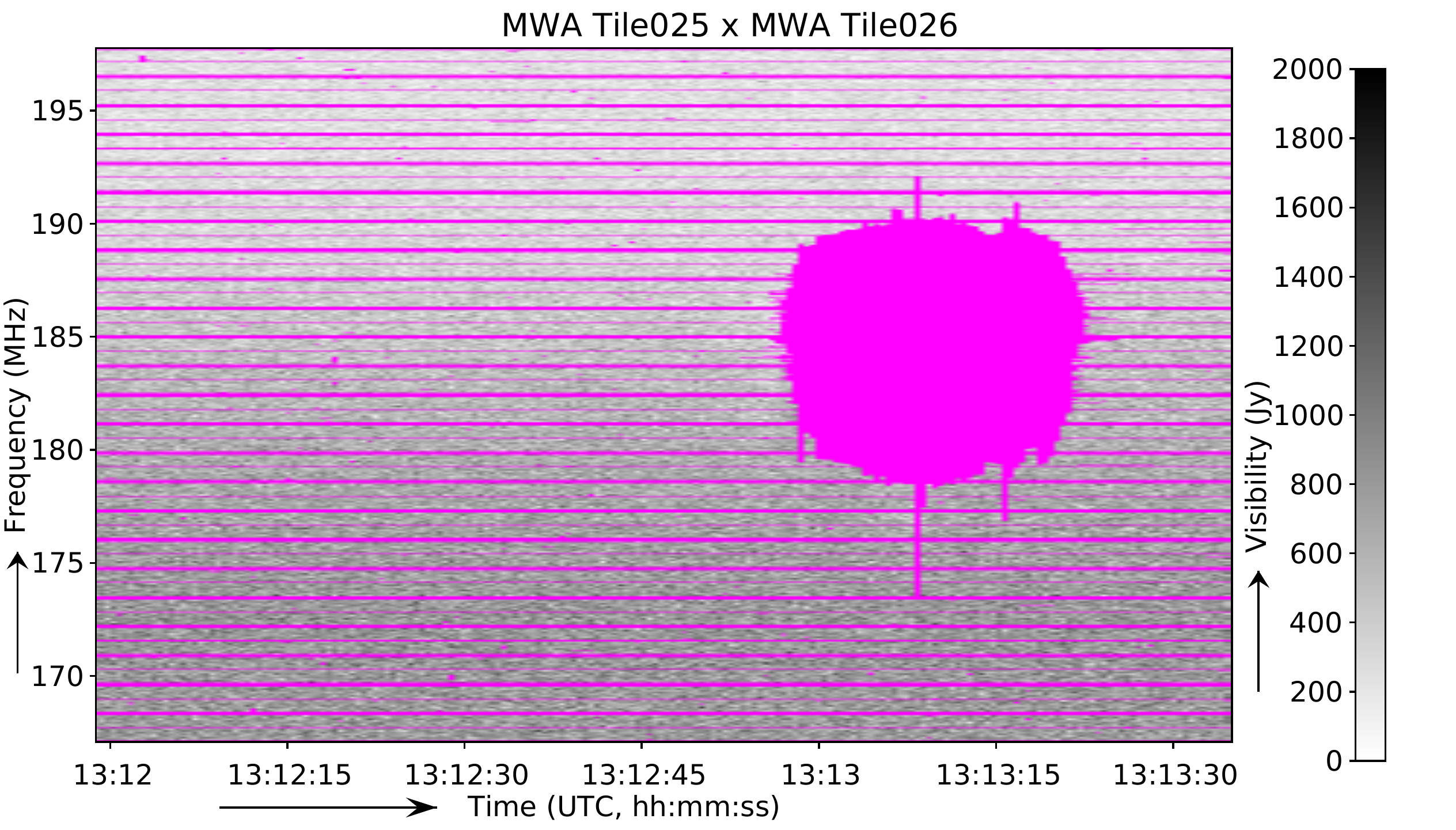}
\end{center}
\caption{DTV RFI briefly visible in an EoR high-band observation. The plots show the Stokes~I amplitudes for a single correlation. The right plot shows the result of RFI detection and invalid channels marked in purple.}
\label{fig:dvb-burst}
\end{figure*}

\noindent\begin{figure*}%
\begin{center}\hspace*{-0.2cm}\includegraphics[width=8cm]{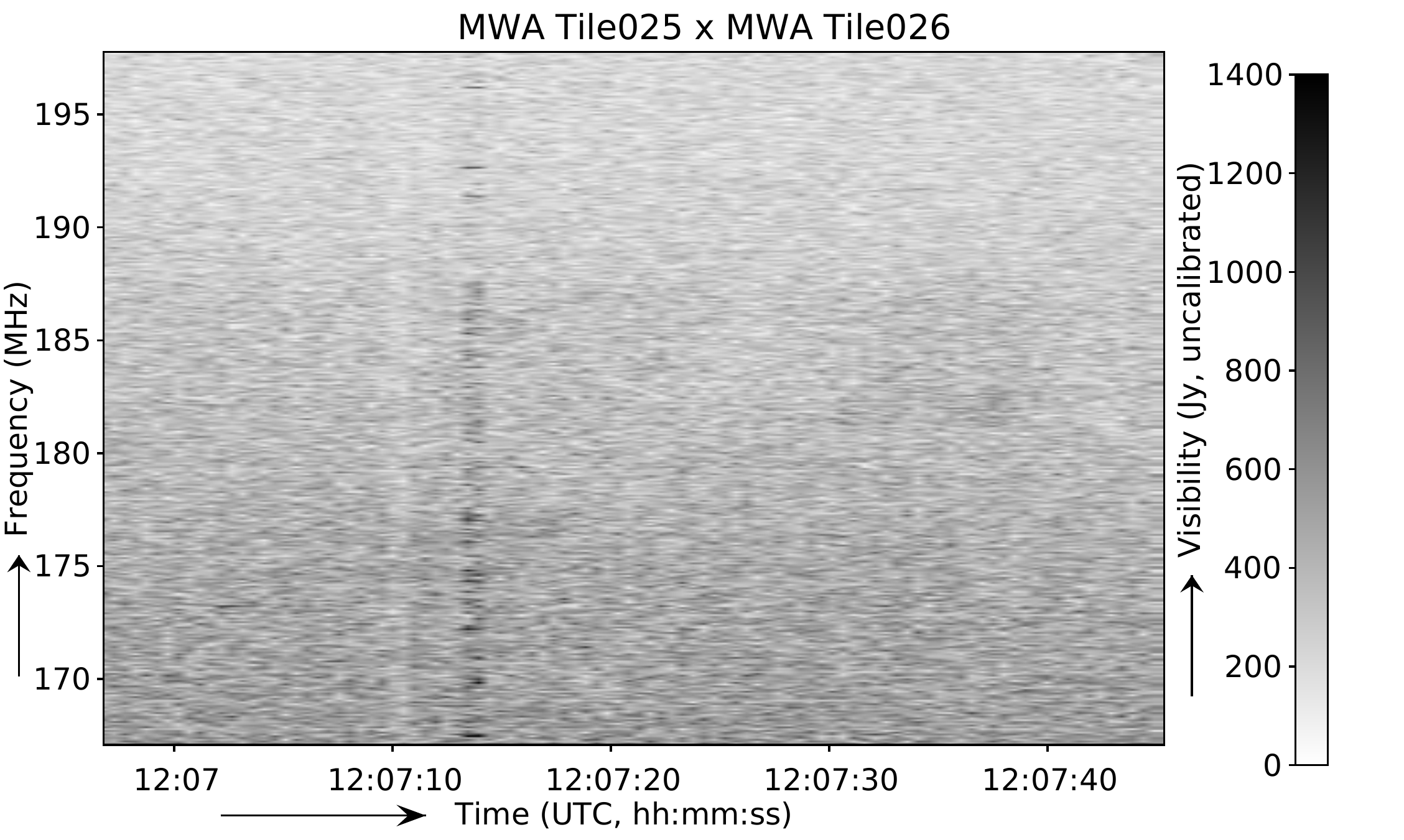}\includegraphics[width=8cm]{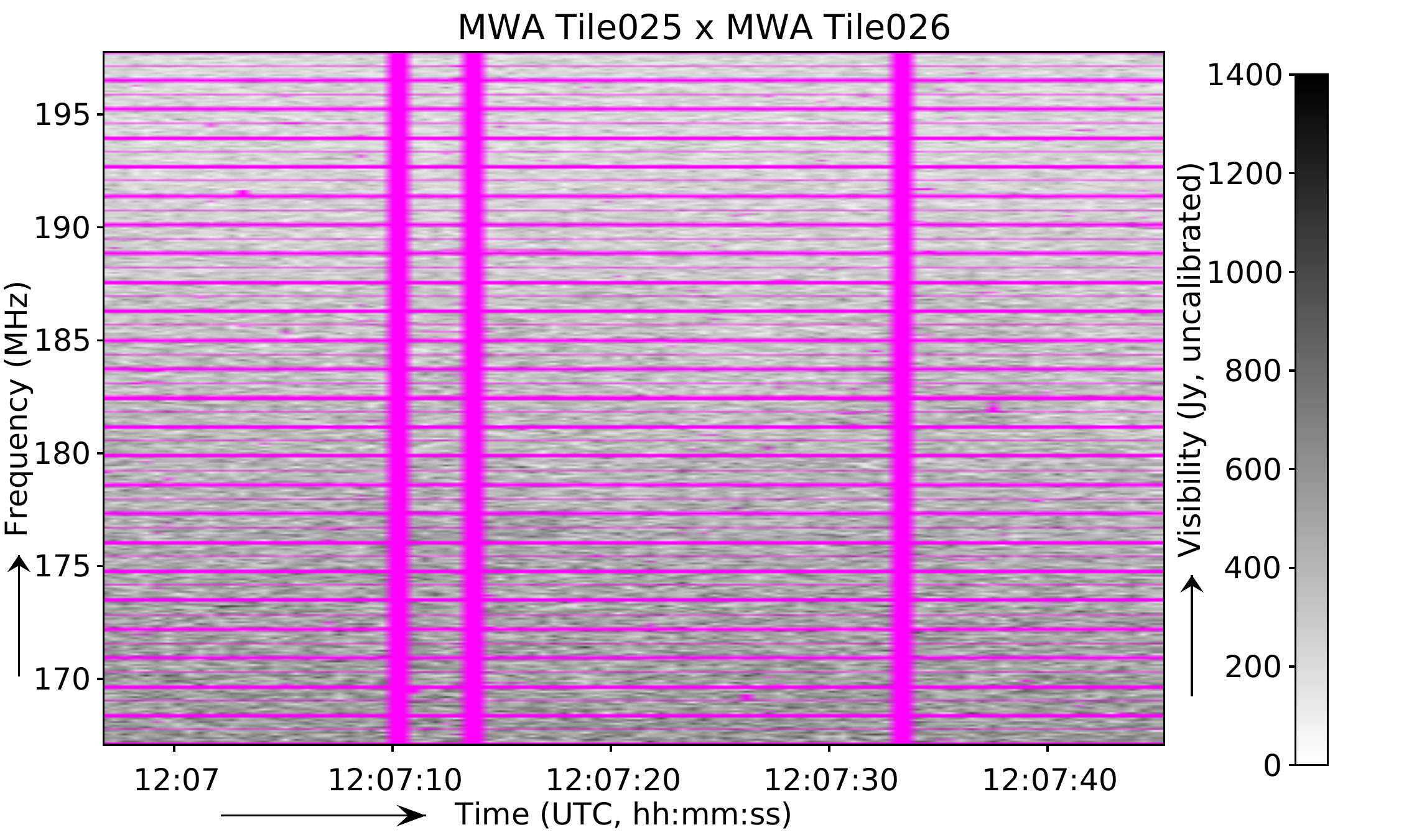}
\end{center}
\caption{Broadband pulses found in an EoR high-band observation. The two panels show Stokes~I visibility amplitudes of a single correlation. In the right plot, the RFI detection flags and invalid channels are marked with purple. Each event occupies 2 timesteps of 0.5~s.}
\label{fig:broadband-pulse}
\end{figure*}

\noindent\begin{figure*}%
\begin{center}
\includegraphics[width=9.1cm]{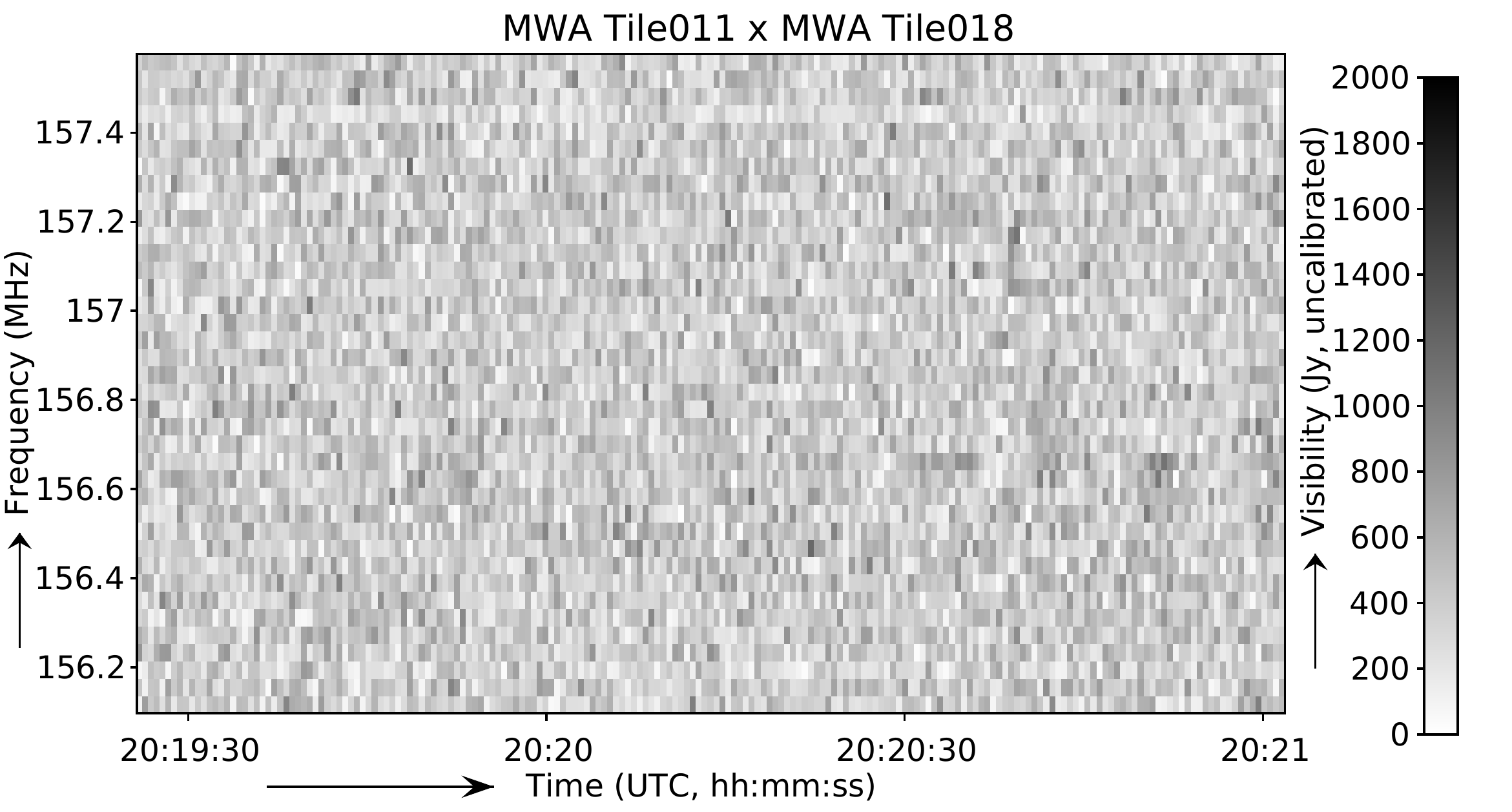}\includegraphics[width=9.1cm]{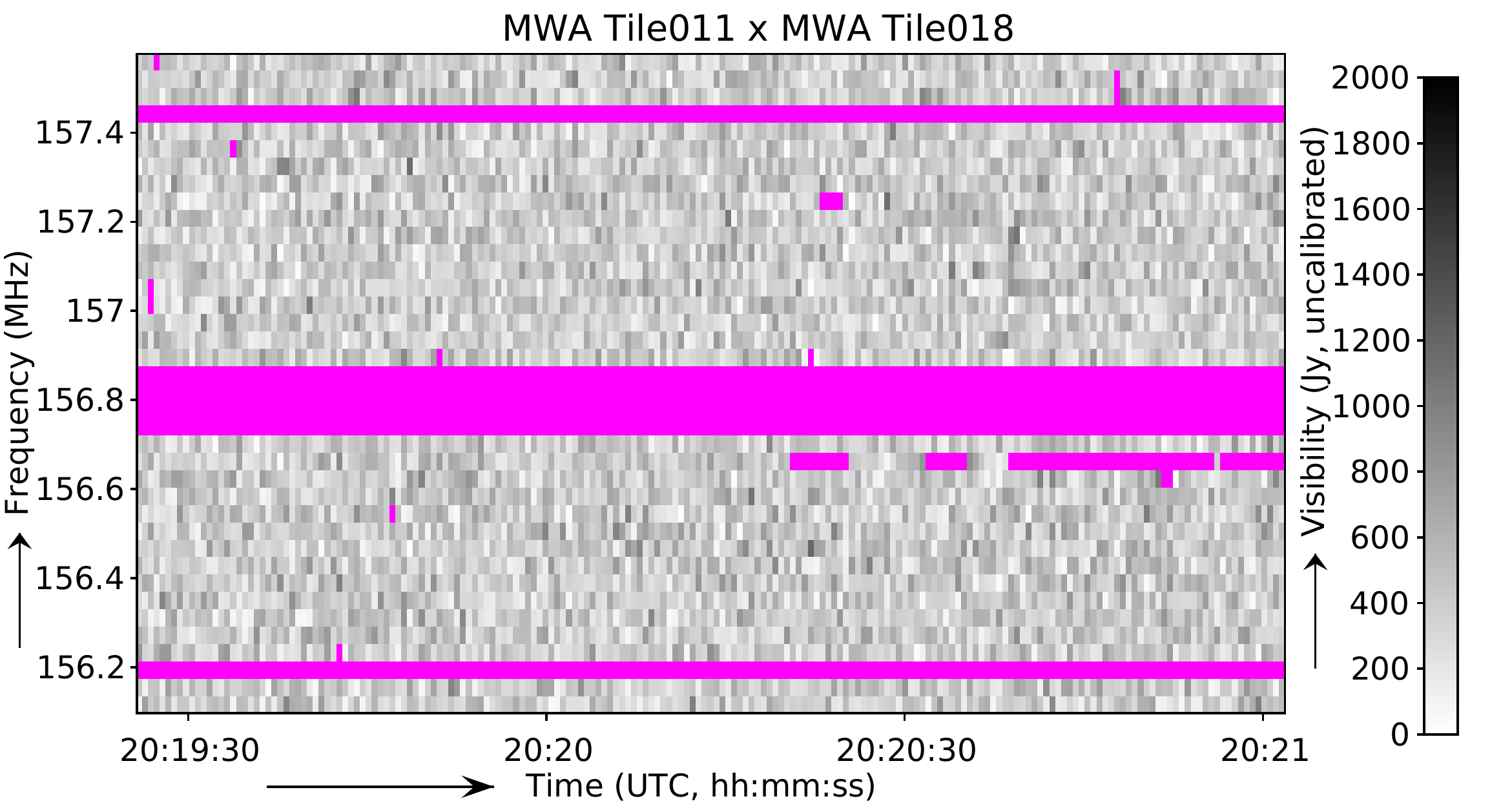}\\
\includegraphics[width=9.1cm]{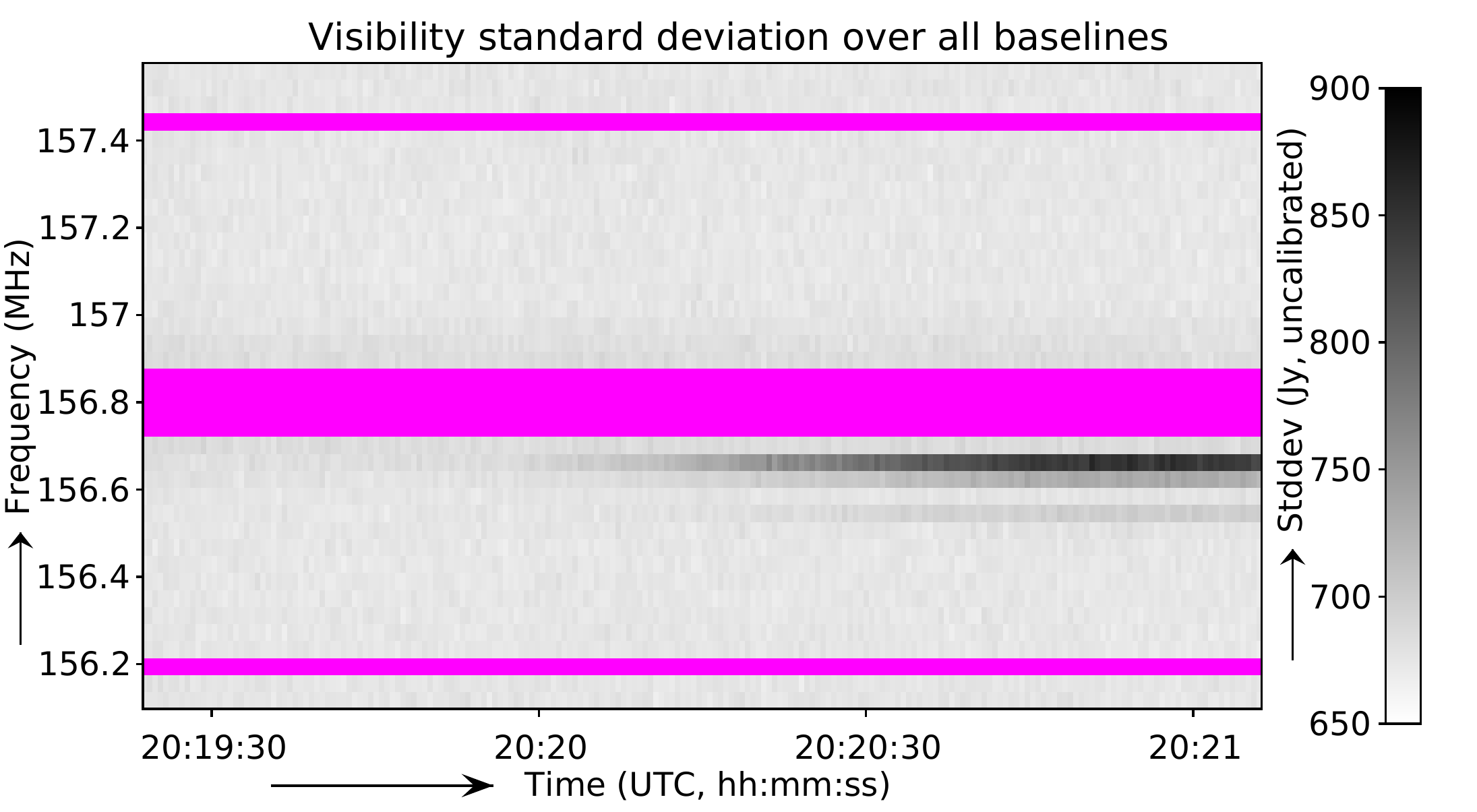}\includegraphics[width=9.1cm]{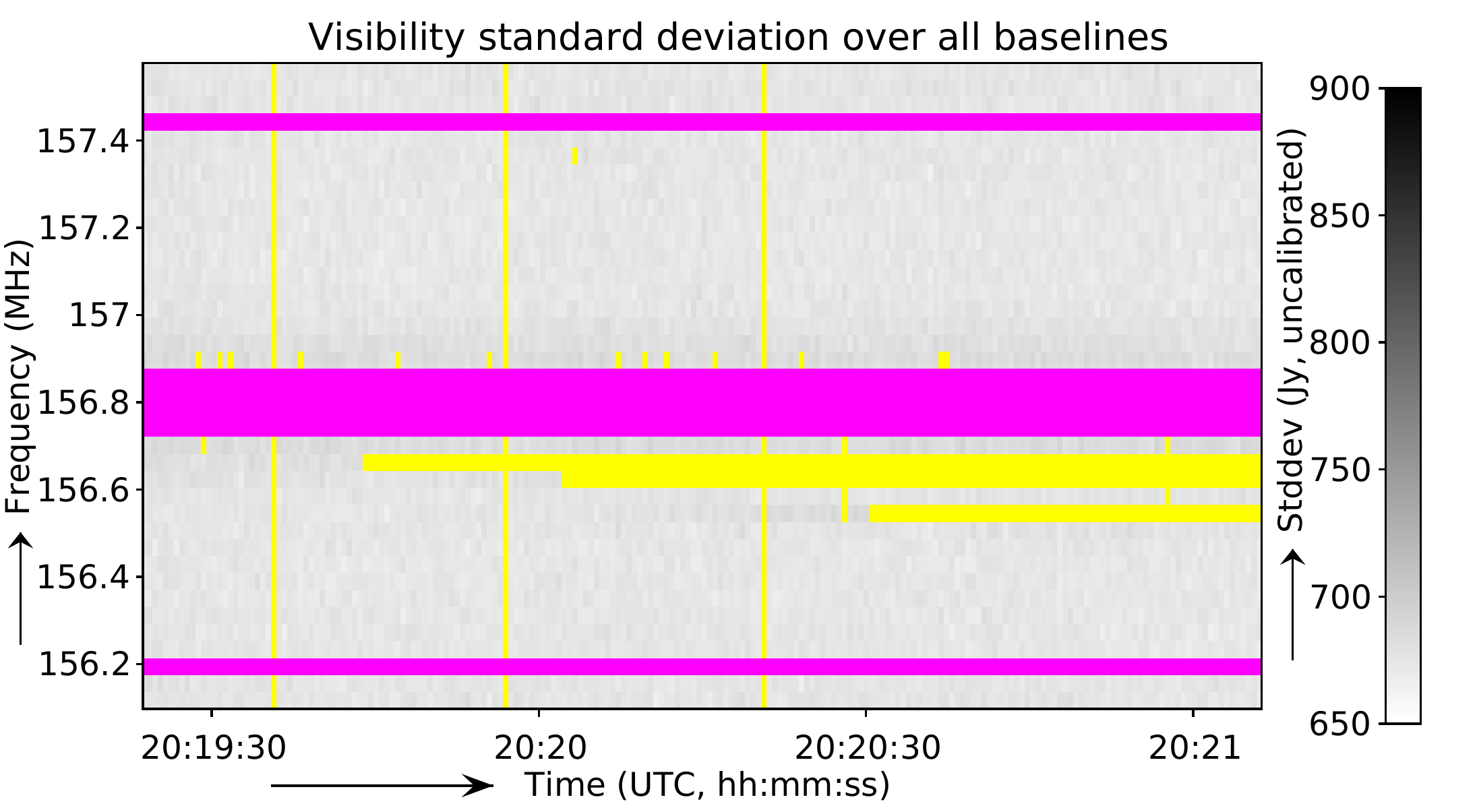}
\caption{A weakly-observed transmitter centred on 156.66 MHz. Top-left window: a single correlation, where the transmitter is barely visible; Top-right window: detected RFI and invalid channels with purple. When performing per-baseline flagging, the transmitter is partly detected; Bottom-left window: when combining data from all baselines, the transmitter becomes clearly visible; Bottom-right window: in yellow, result of executing \textsc{aoflagger} on the standard deviations calculated over all baselines.}\label{fig:156_7}
\end{center}
\end{figure*}

\subsection{RFI types}
A variety of RFI events are observed in the data sets. While there are too many transmitters to show examples for each, it is helpful to understand what kind of events are visible and how they are flagged by the \textsc{aoflagger}. Therefore, a few typical examples are shown.

Fig.~\ref{fig:2m-amateur-band} shows two examples of RFI events in the same EoR low-band snapshot: the top panels display the Stokes~I values of a single correlation, in which a transmitter has been observed in the 2-m amateur band (146 MHz). This is the worst example of contamination in our data sets by this transmitter, and there are snapshots in which the transmitter is not visible at all. This variability might be caused by intrinsic variation, movement of the transmitter or varying propagation conditions. While a change in pointing can also change the appearance of the transmitter, the beam does not change within a single snapshot, while in Fig.~\ref{fig:2m-amateur-band} the strength and affected bandwidth of the transmitter does change during the snapshot. In the bottom panel of Fig.~\ref{fig:2m-amateur-band}, the same snapshot is shown but is zoomed in on a briefly-observed RFI event at 150.17 MHz. This event occupies only a single 40~kHz channel for 2 seconds, and thus is an event which requires flagging at approximately the observed time and frequency resolution or higher for accurate detection.

Besides persistent transmissions that occupy a few channels, transient broad-band events are observed as well. Occasionally, DTV RFI is visible for a brief moment, as for example visible in Fig.~\ref{fig:dvb-burst}. As can be seen in the right-hand plot of Fig.~\ref{fig:dvb-burst}, such a brief interference event is well flagged by the \textsc{aoflagger}. Another transient broad-band example is displayed in Fig.~\ref{fig:broadband-pulse}, which shows strong broad-band pulses of a second in length. Strong pulses such as these are rare and well flagged, but a few weak broad-band pulses are observed in almost every 2-min snapshot. These weak pulses are not visible nor detected in a single baseline correlation, but can be seen in a dynamic spectrum when the power on all baselines is added together. We currently do not know what their origin is. The MWA Voltage Capture System (VCS; \citealt{mwa-voltage-capture-2014}), which allows high time resolution observations, might help to analyse these signals.

Fig.~\ref{fig:156_7} shows a longer RFI contamination at 156.66~MHz that is hardly visible in a single correlation. The \textsc{aoflagger} detects this RFI partially (Fig.~\ref{fig:156_7}, top-right plot), but when plotting the standard deviation over all correlations in a dynamic spectrum, it is evident that this RFI event extends in time beyond what is detected (Fig.~\ref{fig:156_7}, bottom-left plot). The detection becomes more complete when the \textsc{aoflagger} is executed on the standard deviations over all baselines (Fig.~\ref{fig:156_7}, bottom-right plot). This kind of detection is currently not implemented in \textsc{cotter}.

\subsection{Computational performance of \textsc{cotter}}
Because \textsc{cotter} processes the data at high time and frequency resolution, its computational performance is an important consideration. A major contribution to the runtime is the reading and writing of the data, and the runtime is thus influenced by the input-output (IO) disk performance of the host system. Excluding IO, the main computational burden of \textsc{cotter} consists of running \textsc{aoflagger} on the data and collecting the statistics; performing all its other tasks such as applying the cable delays and averaging increases the runtime by approximately $1\%$. To time \textsc{cotter}, we use a high-end desktop computer with 32 GB of memory and a 3.20-GHz Intel Core i7-3930K processor with six cores, and with a 5-disc RAID5 setup. The wall-clock runtime for processing a single 2~min snapshot of 50~GB with a 0.5~s / 40~kHz input resolution, using common averaging settings to output at 2~s / 80~kHz resolution, is split up as follows: 3~min are spent on reading the data, 5~min on RFI detection, and 3.5~min on writing the data. Real-time processing can therefore be achieved by using 6 of such nodes in parallel. Assuming a 138~GFLOPS (Giga-floating-point operations per second) performance of the host computer, the RFI detection requires 25 FLOP/visibility. When expressed as visibilities per time unit, the computational performance of the flagger is independent of the frequency resolution, time resolution and number of antennas. This number can therefore be extrapolated to other telescopes, although the performance of a pipeline which incorporates RFI detection will be strongly dependent on available IO performance, memory bandwidth and other system properties.

\noindent\begin{figure}
\begin{center}\hspace*{-0.2cm}\includegraphics[width=9cm]{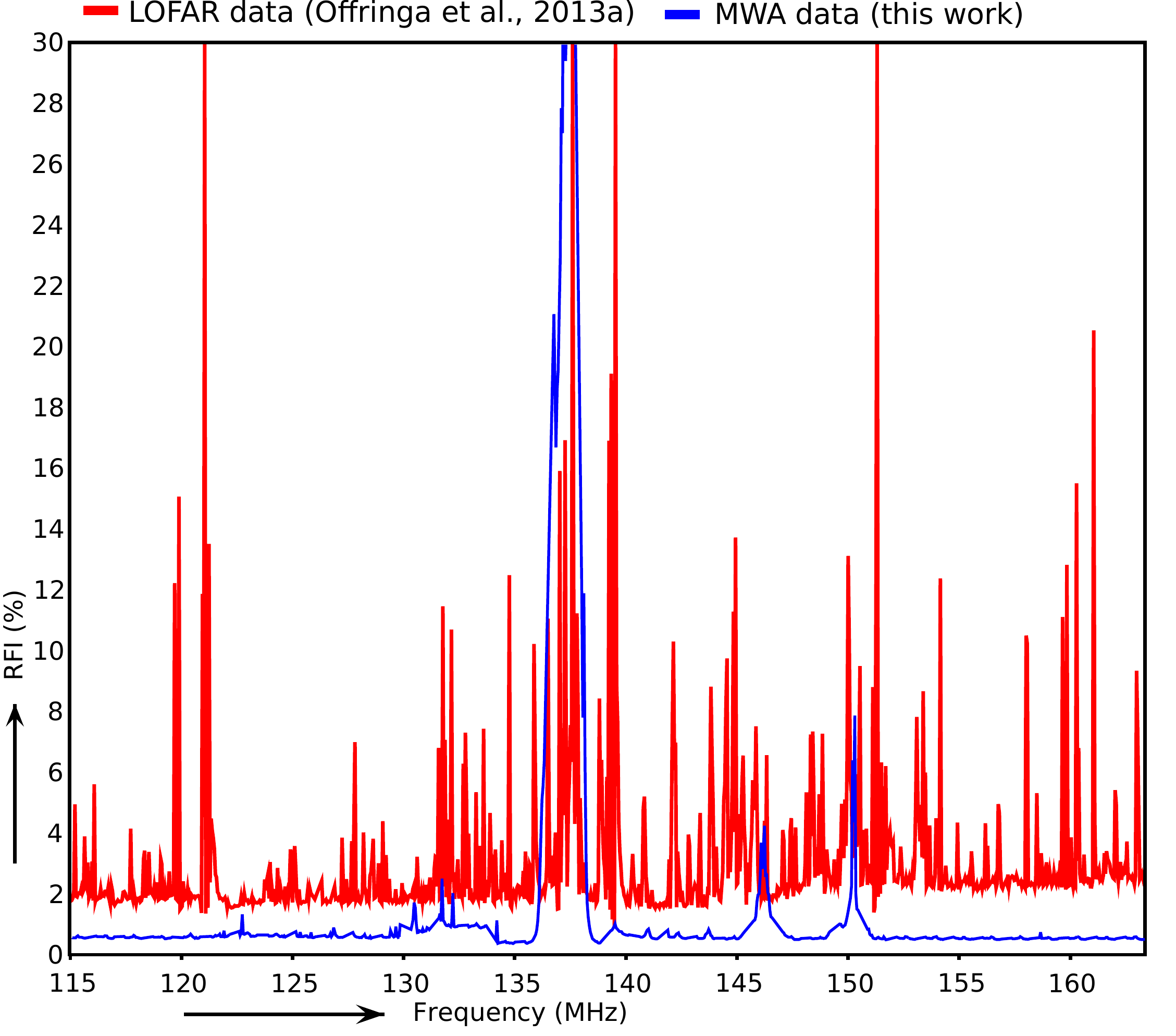}
\caption{Comparison between LOFAR and MWA RFI occupancies. The statistics are resampled to the same frequency resolution of 48~kHz. }
\label{fig:lofar-mwa-occupancy}
\end{center}
\end{figure}

\section{Comparison with LOFAR} \label{ch:lofar-comparison}
The frequency range of the LOFAR high-band antennas (HBAs) overlaps with the MWA frequency range, and this therefore allows a comparison of RFI at the same frequencies between the telescopes. The RFI occupancies for LOFAR from \citet{lofar-radio-environment} and for the MWA from this work are plotted over the HBA frequency sub-range of 115--163~MHz in Fig.~\ref{fig:lofar-mwa-occupancy}. The statistics are regridded to a common frequency resolution of 48~kHz. The average RFI occupancy in this frequency range is 1.65\% for the MWA, while \citet{lofar-radio-environment} reports a 3.18\% occupancy for LOFAR.

It should be noted that the RFI detection for LOFAR was performed at 0.78~kHz while for the MWA it is performed at 40~kHz. As shown by \citet{lofar-radio-environment}, LOFAR observes many RFI events that are only a single or a few 0.78~kHz channels wide, and consequently due to MWA's lower frequency resolution, the MWA will only detect the brightest of such transmissions. \citet{lofar-radio-environment} also show that ionospheric scintillation of Cassiopeia~A triggers detection events, and this is one of the reasons for LOFAR's relatively high minimum occupancy level of $\sim2\%$ RFI, in comparison to 0.5\% for the MWA. In the MWA data, no false detections have been seen that are caused by ionospheric scintillation, likely because of the absence of the strongest sources; Cassiopeia~A and Cygnus~A are only visible at low elevations. An additional explanation could also be that the sidelobe behaviour of the tiled-array beam is also different between MWA and LOFAR due to the difference between MWA tiles and LOFAR stations.

Because of the differences in resolution and the beam forming, it is hard to compare the LOFAR and MWA environments based on detected RFI statistics. Nevertheless, for both telescopes, the automatic detection strategies have been optimized such that as little data as possible are thrown away, but to be sufficient for further data reduction. These values can therefore be interpreted and compared as being the minimum loss of data due to RFI.

After RFI detection and excision, power spectra from both MWA and LOFAR show smooth curves without artefacts that can be attributed to leakage, and RFI does not lead to detectable artifacts in the resulting image at the thermal noise limit. From the difference between 1.65\% or 3.18\% loss of data at the MWA and LOFAR sites respectively, it is clear that the impact of RFI is smaller for the MWA, but RFI mitigation is still required with similar per-sample computational requirements as for LOFAR. In both cases, RFI occupancy levels are small and RFI flagging is effective. Nevertheless, a benefit of MWA's remote radio-quiet site is that it allows observations in the 88--108~MHz FM-station and 174-230 MHz DTV bands. Initial MWA experiments have confirmed the availability of these bands for science \citep{mckinley-moon-2013,hurley-walker-mwacs-2014}. Moreover, for both LOFAR and MWA, a primary science driver is the detection of signals from the EoR. The required integration time for such a detection of approximately a hundred nights has currently not been achieved, and it could be that RFI becomes problematic when reaching lower noise levels. Initial results with LOFAR predict that RFI will not prevent such a detection \citep{ncp-eor-yatawatta, offringa-rfi-distributions}.

\section{Conclusions \& discussion} \label{ch:conclusions-and-discussion}
We have described the automated RFI detection strategy for the MWA and shown RFI statistics at various frequencies and over various nights. After RFI detection and excision, data can be calibrated and imaged without artefacts visible at the thermal or confusion noise level, with the exception of the ORBCOMM bands at 137~MHz. Also, DTV signals are seen $\sim3\%$ of the time, and when present make observing in the 174--195~MHz range impossible. Over the full GLEAM range of 72--231~MHz, 1.1\% of the data are detected and flagged by the \textsc{aoflagger}, and these RFI events are attributed to several different transmitters. Some residual RFI is seen in the FM-station bands, but these frequencies are usable after our described automated RFI detection. The issue of RFI has become smaller by building at a radio-quiet site, but still requires adequate mitigation. When observing continuously with the MWA, a few fast computing nodes are permanently required for real-time RFI detection.

SKA-low will be built at the same location as the MWA, and hence several lessons can be learned. First of all, it is clear that the SKA will need to be able to handle some amount of RFI. This requires computational power to perform the RFI detection, and a receiver signal path with headroom sufficient to avoid gain compression by RFI. Second, \textsc{cotter} relies on the fact that a single computing node can still hold a reasonable amount of MWA data in memory for the RFI detection, but because of SKA's large number of elements and high time and frequency resolution, this will be a more challenging problem. Thirdly, flagging MWA data is slightly complicated due to the first poly-phase filter and digital gains. The extra per-subband gain correction that is required for the MWA makes the detection less stable, and the sub-band bandpass makes it harder to recognize RFI patterns in frequency direction. Therefore, for accurate RFI detection it is best to have a smooth response over a large instantaneous bandwidth. Finally, the presence of short and spectrally-narrow RFI events confirms that detection at high time and frequency resolution improves accuracy.

The \textsc{aoflagger} RFI detection strategy, originally developed for LOFAR, works well for the MWA. Faint RFI events such as the ones in Figs.~\ref{fig:broadband-pulse} and \ref{fig:156_7}, or complex events such as the one in Fig.~\ref{fig:dvb-burst}, are not adequately detected by a single-sample thresholding algorithm, but \textsc{aoflagger}'s SumThreshold and SIR-operator algorithms are able to flag such events. These algorithms have gained some popularity; besides the use of \textsc{aoflagger} by individual astronomers, \textsc{miriad}'s PGFLAG task implements both the SumThreshold and SIR-operator algorithms, and the pipeline for eMERLIN \citep{serpent-peck-2013} implements the SumThreshold algorithm. Nevertheless, other projects still use single-sample thresholding, e.g. PAPER \citep{parsons-paper-eorlimit-2014}. Since strong RFI is seen practically everywhere, and because the apparent strength of RFI events will follow a power-law distribution, many faint transmitters will interfere with observations of any (terrestrial) radio observatory. Using algorithms with low sensitivity will detect fewer of these, and thus result in RFI becoming more quickly a problem in deep integration projects.

The current sensitivity of \textsc{aoflagger} is enough for calibration and imaging of MWA data. However, \textsc{aoflagger} does not perform well on continuous broadband RFI such as DTV, which is occasionally present due to tropospheric ducting or ionospheric activity. To remove DTV, a second detection round is required, and the current methodology to handle DTV RFI is to delete all affected snapshots, or even the entire night. This requires hardly any computational power, because the required visibility statistics are collected in \textsc{cotter}. For deep-integration projects, such as the EoR projects, it might be that low-level RFI will show up at lower noise levels. One way to increase the sensitivity of the flagger would be to operate on the summed power of multiple baselines. Fig.~\ref{fig:156_7} shows that this increases the detectability of certain RFI events significantly.

Projects that try to detect very weak signals, such as the EoR projects, have to be careful that a (non)detection has not been affected by RFI leakage. One consideration is the storage of data: the symptoms of leaked RFI are harder to identify once visibilities have been averaged or compression techniques such as Delay/Delay-Rate filtering \citep{parsons-paper-eorlimit-2014} have been applied. Storing more information allows better verification of a detection and, if necessary, increases the chance of successfully performing further RFI excision.

\section*{Acknowledgements}
This scientific work makes use of the NCI National Facility in Canberra, Australia, which is supported by the Australian Commonwealth Government.

This scientific work makes use of the Murchison Radio-astronomy Observatory, operated by CSIRO. We acknowledge the Wajarri Yamatji people as the traditional owners of the Observatory site. Support for the MWA comes from the U.S. National Science Foundation (grants AST-0457585, PHY-0835713, CAREER-0847753, and AST-0908884), the Australian Research Council (LIEF grants LE0775621 and LE0882938), the U.S. Air Force Office of Scientific Research (grant FA9550-0510247), and the Centre for All-sky Astrophysics (an Australian Research Council Centre of Excellence funded by grant CE110001020). Support is also provided by the Smithsonian Astrophysical Observatory, the MIT School of Science, the Raman Research Institute, the Australian National University, and the Victoria University of Wellington (via grant MED-E1799 from the New Zealand Ministry of Economic Development and an IBM Shared University Research Grant). The Australian Federal government provides additional support via the Commonwealth Scientific and Industrial Research Organisation (CSIRO), National Collaborative Research Infrastructure Strategy, Education Investment Fund, and the Australia India Strategic Research Fund, and Astronomy Australia Limited, under contract to Curtin University. We acknowledge the iVEC Petabyte Data Store, the Initiative in Innovative Computing and the CUDA Center for Excellence sponsored by NVIDIA at Harvard University, and the International Centre for Radio Astronomy Research (ICRAR), a Joint Venture of Curtin University and The University of Western Australia, funded by the Western Australian State government. 

\DeclareRobustCommand{\TUSSEN}[3]{#3}

\bibliography{references}

\label{lastpage}

\end{document}